\documentclass[twocolumn,superscriptaddress,floatfix,preprintnumbers, nofootinbib,hyperref]{revtex4-2} 
\pdfoutput=1
\usepackage[colorlinks=true,breaklinks=true]{hyperref}
\usepackage[normalem]{ulem}
\usepackage{slashed}
\usepackage[utf8]{inputenc}
\hypersetup{allcolors=[rgb]{0.0 0.0 0.6},linkcolor=[rgb]{0.75 0.05 0.05}}
\usepackage{amsmath,amssymb}
\usepackage{epsfig}  
\usepackage{graphicx}   
\usepackage{slashed}       
\usepackage{url}
\usepackage{color}
\usepackage{multirow}
\usepackage{orcidlink}
\usepackage{comment}
\usepackage{amssymb}

\usepackage{tikz}
\usepackage{tikz-feynman}
\usetikzlibrary{matrix}
\usetikzlibrary{matrix, positioning}
\tikzfeynmanset{compat=1.1.0}

\usepackage{empheq,etoolbox}
\patchcmd{\subequations}% <cmd>
  {\theparentequation\alph{equation}}% <search>
  {\theparentequation.\alph{equation}}% <replace>
  {}{}% <success><failure>
  
\newcommand{\bk}{{\bf k}}

\newcommand{\bq}{{\bf q}}
\newcommand{\bp}{{\bf p}}
\newcommand{\bP}{{\bf P}}

\newcommand{\beq}{\begin{equation}}
\newcommand{\eeq}{\end{equation}}

\long\def\exclude#1{}

\newcommand{\tP}{{|\tilde{\bP}|}}
\newcommand{\ttP}{{\tilde{P}}}
\newcommand{\ttp}{{\tilde{p}}}
\newcommand{\ttq}{{\tilde{q}}}
\newcommand{\tq}{{|\tilde{\bq}|}}
\newcommand{\tx}{{\tilde{x}}}

\allowdisplaybreaks

\bibpunct{[}{]}{,}{n}{}{,}

\begin{document}

\title{Millicharged Particle Production During Late-Stage Stellar Evolution
}

\author{Damiano F.\ G.\ Fiorillo \orcidlink{0000-0003-4927-9850}}
\email{damianofg@gmail.com}
\affiliation{Istituto Nazionale di Fisica Nucleare (INFN), Sezione di Napoli, Complesso Universitario di Monte Sant'Angelo, Via Cintia, 80126 Napoli, Italy}
\affiliation{Deutsches Elektronen-Synchrotron DESY,
Platanenallee 6, 15738 Zeuthen, Germany}

\author{Giuseppe Lucente
\orcidlink{0000-0003-1530-4851}}
\email{lucenteg@slac.stanford.edu}
\affiliation{SLAC National Accelerator Laboratory, 2575 Sand Hill Rd, Menlo Park, CA 94025}

\author{Jeremy Sakstein
\orcidlink{0000-0002-9780-0922}}
\email{sakstein@hawaii.edu}
\affiliation{Department of Physics \& Astronomy, University of Hawai'i, Watanabe Hall, 2505 Correa Road, Honolulu, HI, 96822, USA}

\author{Edoardo Vitagliano
\orcidlink{0000-0001-7847-1281}}
\email{edoardo.vitagliano@unipd.it}
\affiliation{Dipartimento di Fisica e Astronomia, Università degli Studi di Padova,
Via Marzolo 8, 35131 Padova, Italy}
\affiliation{Istituto Nazionale di Fisica Nucleare (INFN), Sezione di Padova,
Via Marzolo 8, 35131 Padova, Italy}

\smallskip

\begin{abstract}
Stars are natural sources of feebly interacting particles, including putative particles with mass $m_\chi$ and electric charge $qe$. The emission of such millicharged particles (MCPs) causes an energy loss which can alter stellar evolution. While MCP production rates have been computed for different plasma parameters, they have yet to be derived for the conditions relevant to late stages of stellar evolution, in which the temperature can reach values $T\simeq 10-100\,\rm keV$ while the plasma frequency is $\omega_{\rm pl}\ll T$. In this paper, we compute the MCP energy-loss rates  relevant for pre-supernova objects, finding three different regimes in which the dominant processes are respectively plasmon decay ($m_\chi< \omega_{\rm pl}/2$), Compton-like scattering ($m_\chi> \omega_{\rm pl}/2$, $T\lesssim 0.5\,\rm MeV$), and electron-positron annihilation. We obtain semi-analytical fits for the energy-loss rates suitable for  implementation in stellar evolution codes.

\end{abstract}

\maketitle

\tableofcontents

\section{Introduction}
While charge quantization seems to be supported by well-established observations, physics beyond the Standard Model (SM) must be evoked to enforce it~\cite{Foot:1990mn}. Therefore, one could take the opposite point of view: that charge quantization is after all violated by the putative existence of millicharged particles (MCPs) $\chi$, whose interactions are described by
\begin{equation}
\mathcal{L} \supset q e\,\overline{\chi} \gamma^\mu\chi\,A_\mu + \overline{\chi}(i\slashed{\partial}-m_\chi)\chi \,.
\end{equation}
There are several ways in which such particles might arise, e.g. from an $SU(3)\times SU(2)$ singlet with hypercharge $Y=2q$~\cite{Mohapatra:1990vq}; alternatively, neutrinos might carry a millicharge (see e.g. Ref.~\cite{Giunti:2014ixa} and references therein). Finally, a widely studied case is a novel $U(1)$ massless gauge boson, the dark photon, mixing kinetically with the SM photon at low energies~\cite{Holdom:1985ag}; a dark fermion to this boson would naturally develop a millicharge, even though the true $U(1)$ charge remains quantized.

Similarly to other putative particles, MCPs would be abundantly produced in stellar cores, altering the standard picture of stellar evolution~\cite{Raffelt:1996wa} at various stages, including main-sequence stars and the Sun~\cite{Vinyoles:2015khy}, red giants~\cite{Dobroliubov:1989mr,Davidson:1991si,Babu:1992sw,Haft:1993jt,Davidson:1993sj,Davidson:2000hf,Fung:2023euv}, horizontal-branch stars~\cite{Raffelt:1996wa,Davidson:2000hf,Vogel:2013raa}, white dwarfs~\cite{Dobroliubov:1989mr,Davidson:1991si,Babu:1992sw,Davidson:1993sj,Davidson:2000hf},
and
supernovae~\cite{Mohapatra:1990vq,Davidson:1993sj,Davidson:2000hf,Chang:2018rso,Fiorillo:2024upk}. 

 \begin{figure*}[t!]
    \centering
    \includegraphics[width=0.49\textwidth]{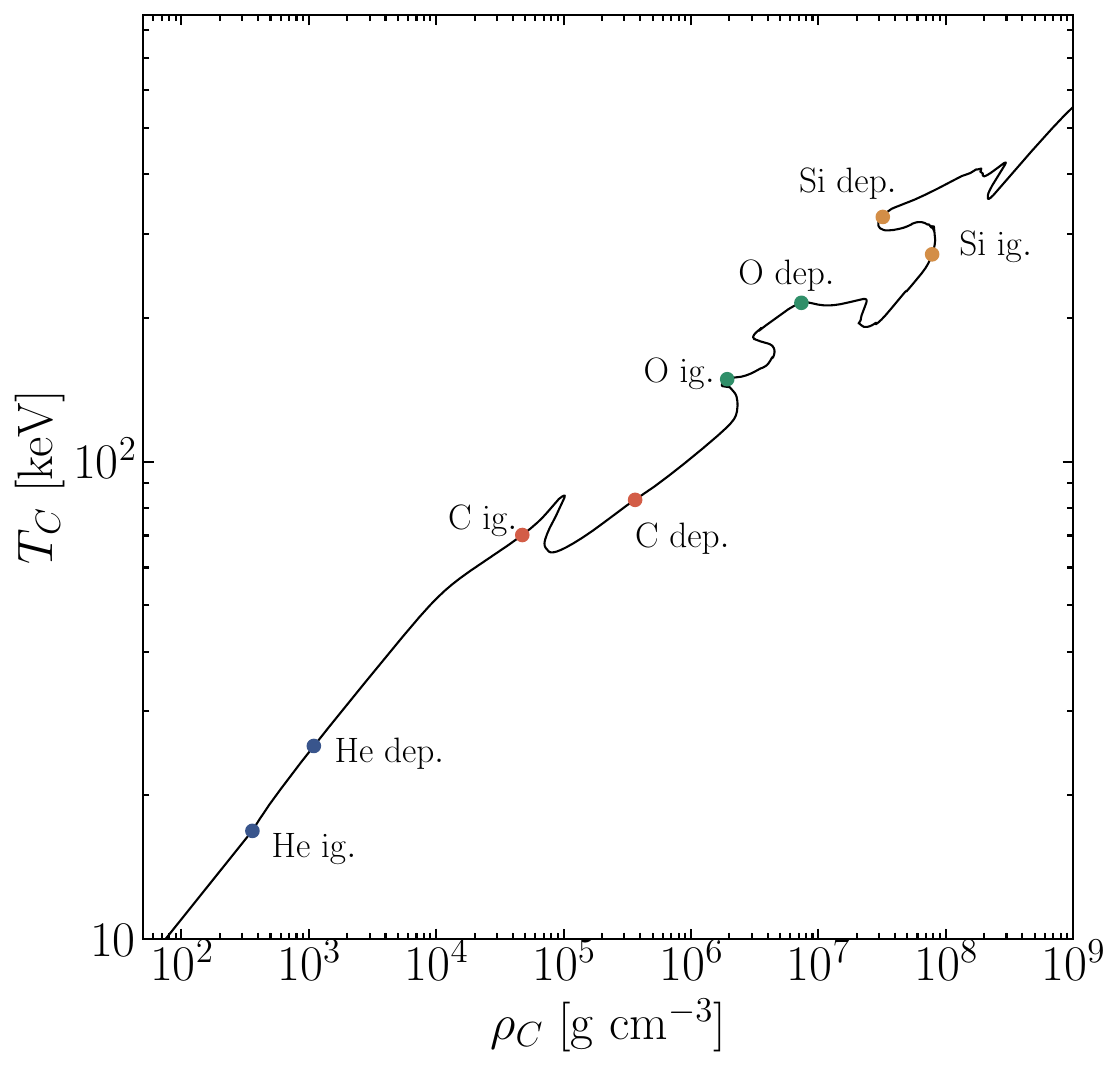}
    \includegraphics[width=0.49\textwidth]{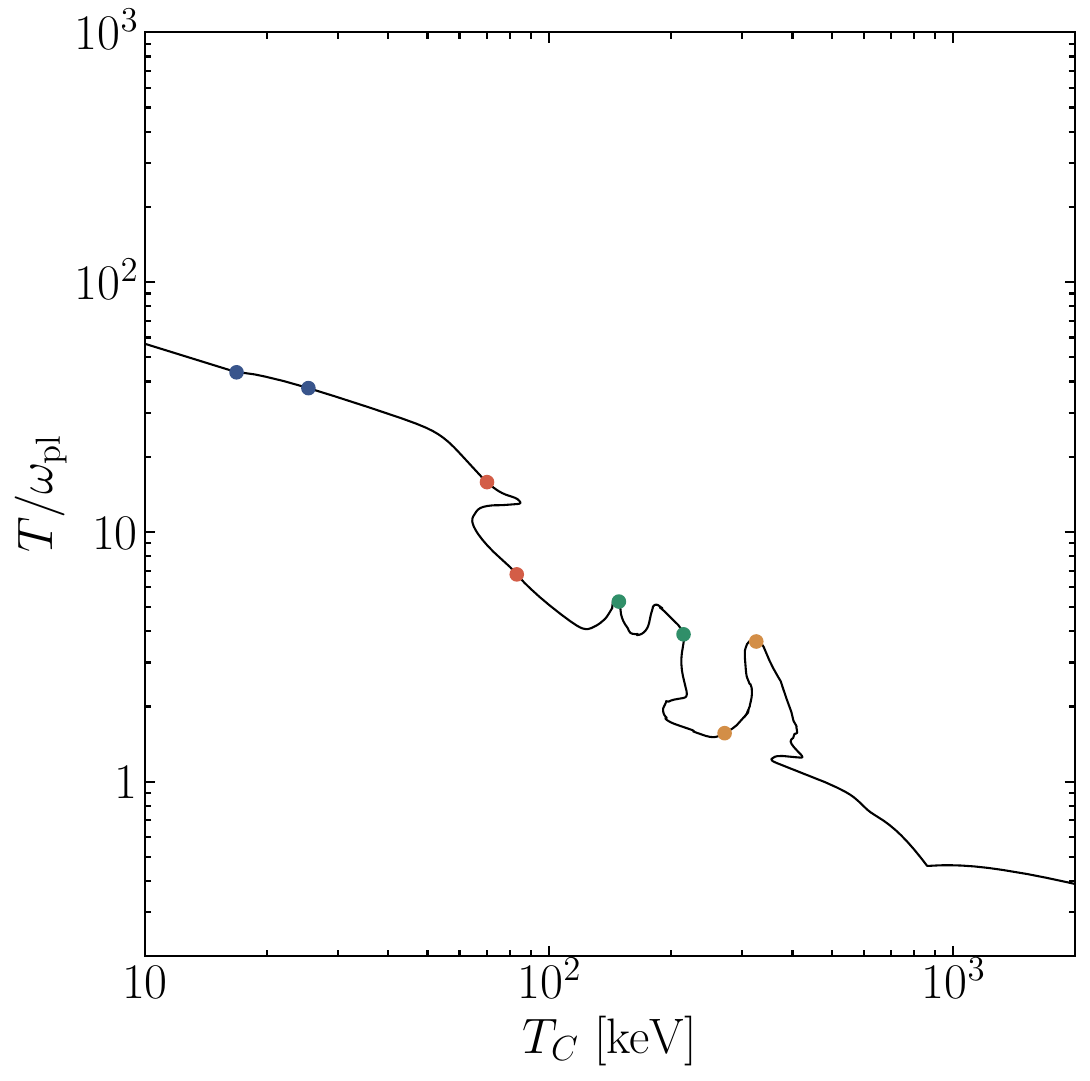}
    \caption{Left panel: Evolutionary track of a $20\,M_\odot$ stellar model in the $T_C\,–\,\rho_C$ plane. Colored markers indicate key evolutionary stages. Right panel: Evolution of the ratio $T/\omega_{\rm pl}$ for the same $20\,M_\odot$ model.}
    \label{fig:ev_track}
\end{figure*}

In this work, we are primarily interested in the production of MCPs in massive, late-stage stellar interiors. Earlier-stage stellar cores, which are significantly colder, are best suited to probe the emission of MCPs with masses below a few keV; in this respect, red giants offer the most sensitive probe~\cite{Fung:2023euv}, though both the Sun~\cite{Vinyoles:2015khy} and white dwarfs~\cite{Vogel:2013raa} can be competitive laboratories. On the other hand, stellar cores during the supernova phase can reach much larger temperatures, so that one can constrain feebly interacting particles up to masses of hundreds of MeV~\cite{Mohapatra:1990vq,Davidson:2000hf,Chang:2018rso}. However, their production has only recently been studied in detail, including the effects of their self-interactions~\cite{Fiorillo:2024upk}.~As recently shown in the context of radiatively decaying particles~\cite{Candon:2024eah}, the late-stage, pre-supernova conditions provide an optimal regime to constrain keV–MeV feebly interacting particles, motivating a dedicated study of their emission. The energy-loss rates computed here are the foundations for forthcoming work in which we will search for MCP signals in heavy stars.

The paper is structured as follows. Section~\ref{sec:plasma}
reviews the general conditions of the stellar plasma in late-stage stellar evolution. In Sections~\ref{sec:pldec}, \ref{sec:compt}, and \ref{sec:PairProd} we revisit the relevant production processes, respectively plasmon decay ($\gamma_{L,T}\to\chi \overline{\chi}$), Compton-like scattering ($e^-+\gamma\to e^-+\overline{\chi}+\chi$), and electron-positron annihilation ($e^++e^-\to \overline{\chi}+\chi$). Finally, Section~\ref{sec:disc} provides a summary of our findings.

\section{Plasma conditions for MCP production in late-stage stars}\label{sec:plasma}

Depending on their initial mass, stars can reach extremely high temperatures and densities. If their initial mass is $M\gtrsim 8\,M_\odot$, where $M_\odot$ is the solar mass, they will eventually collapse, triggering the explosion of a core-collapse supernova (SN)~\cite{Janka:2006fh,Janka:2012wk} and the release of a large neutrino flux~\cite{Vitagliano:2019yzm,Raffelt:2025wty}. The observations of Kamiokande~II~\cite{Kamiokande-II:1987idp,Hirata:1988ad} and the Irvine-Michigan-Brookhaven~\cite{Bionta:1987qt,IMB:1988suc} experiment of SN~1987A~\cite{Koshiba:1992yb} are compatible with a scenario in which temperatures can get as large as tens of MeV (see Ref.~\cite{Fiorillo:2023frv} for a recent comparison with modern simulations).

Before collapse, such stars can spend millions of years in a post-main-sequence stage, in which the typical temperature in their cores can reach $T\sim 10-100\,\mathrm{keV}$, with densities $\rho\sim 10^3\,\mathrm{g/cm}^3$, yielding an electron density of approximately $n_e\sim 10^{27}\,\mathrm{cm}^{-3}$~\cite{Woosley:2002zz,Langer:2012jz,Laplace:2021vre} (see also the Supplemental Material of Ref.~\cite{Candon:2024eah}). This implies a Fermi momentum for electrons $p_F\sim (3\pi^2 n_e)^{1/3}\sim 50\,\mathrm{keV}$, and in turn a chemical potential $\mu_e\sim p_F^2/2m_e\sim 1\,\mathrm{keV}\ll T$. This implies a non-degenerate system. The typical plasma frequency is $\omega_{\rm pl}=\sqrt{4\pi \alpha n_e/m_e}\sim 1\,\mathrm{keV}$, where $\alpha$ is the fine-structure constant and $m_e$ is the electron mass, so $\omega_{\rm pl}\ll T$. Thus, our conditions of interest are those of a non-relativistic, non-degenerate plasma. We show in the left panel of Fig.~\ref{fig:ev_track} the evolutionary track of a representative $20\,M_\odot$ stellar model in the plane of central temperature $T_C$ versus central density $\rho_C$. The labeled points indicate the main burning stages, from He ignition (ig.) through successive fuel depletion (dep.) phases, up to core collapse after Si depletion. As shown in the right panel of Fig.~\ref{fig:ev_track}, the condition $\omega_{\rm pl}\ll T$ holds during most of the stellar evolution, while $\omega_{\rm pl}$ becomes comparable to $T$ only in the final stages approaching core collapse after Si depletion.

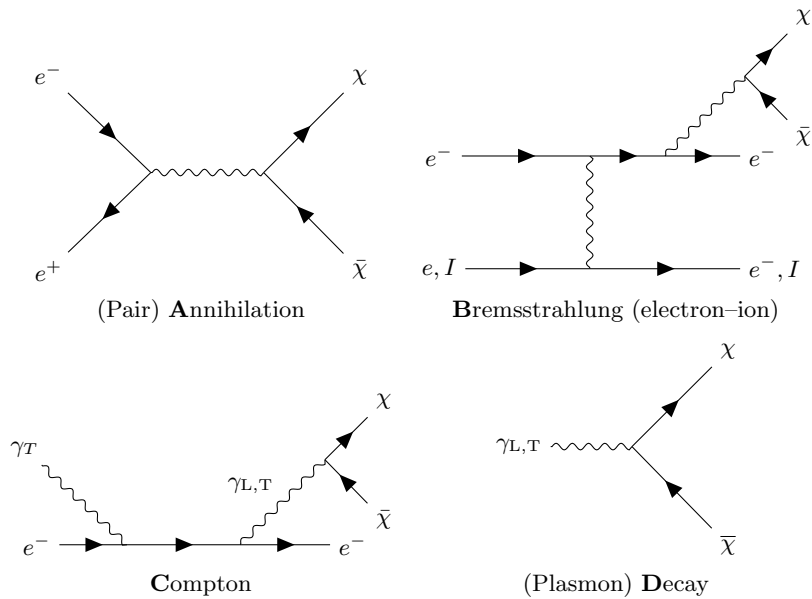
\begin{figure*}[t!]
\centering

\begin{tabular}{cc}

% ================= ROW 1 =================

\begin{minipage}[t]{0.3\textwidth}
\centering
% --- Pair annihilation ---
\begin{tikzpicture}[scale=0.5]
\begin{feynman}
\vertex (a);
\vertex [above left=of a] (e1){\(e^-\)};
\vertex [below left=of a] (e2){\(e^+\)};
\vertex [right=of a] (b);
\vertex [above right=of b] (f1){\(\chi\)};
\vertex [below right=of b] (f2){\(\bar{\chi}\)};
\diagram* {
(e1) -- [fermion] (a) -- [fermion] (e2),
(a) -- [photon] (b),
(f2) -- [fermion] (b) -- [fermion] (f1),
};
\end{feynman}
\end{tikzpicture}

\small (Pair) {\bf{A}}nnihilation
\end{minipage}
&
\begin{minipage}[t]{0.3\textwidth}
\centering
% --- Bremsstrahlung ---
\begin{tikzpicture}[scale=0.5]
\begin{feynman}
\vertex (a){\(e,I\)};
\vertex [right=2 cm of a] (b);
\vertex [right=2 cm of b] (c){\(e^-,I\)};
\vertex [above=of b] (d);
\vertex [left=1.7 cm of d] (e){\(e^-\)};
\vertex [right=1 cm of d] (fp1);
\vertex [right=1 cm of fp1] (f){\(e^-\)};
\vertex [above right= 1.5 cm of fp1] (p1);
\vertex [above right=0.8 cm of p1] (f1){\(\chi\)};
\vertex [below right=0.8 cm of p1] (f2){\(\bar{\chi}\)};
\diagram* {
(a) -- [fermion] (b) -- [fermion] (c),
(b) -- [photon] (d),
(e) -- [fermion] (d) -- [fermion] (fp1),
(fp1) -- [fermion] (f),
(fp1) -- [photon] (p1),
(p1) -- [fermion] (f1),
(f2) -- [fermion] (p1),
};
\end{feynman}
\end{tikzpicture}

\small {\bf{B}}remsstrahlung (electron--ion)
\end{minipage}
\\[0.5cm]

% ================= ROW 2 =================

\begin{minipage}[t]{0.3\textwidth}
\centering
% --- Compton ---
\begin{tikzpicture}[scale=0.5]
\begin{feynman}
\vertex (a) {\(e^-\)};
\vertex [right=1.2 of a] (b);
\vertex [above left=of b] (c) {\(\gamma_T\)};
\vertex [right= of b] (d);
\vertex [right=1.2 of d] (e){\(e^-\)};
\vertex [above right=1.6 cm of d] (p1);
\vertex [above right=0.8 cm of p1] (f1){\(\chi\)};
\vertex [below right=0.8 cm of p1] (f2){\(\bar{\chi}\)};
\diagram* {
(a) -- [fermion] (b) -- [fermion] (d) -- [fermion] (e),
(c) -- [photon] (b),
(d) -- [photon, edge label=\(\gamma_{\rm L,T}\)] (p1),
(p1) -- [fermion] (f1),
(f2) -- [fermion] (p1),
};
\end{feynman}
\end{tikzpicture}

\small {\bf C}ompton
\end{minipage}
&
\begin{minipage}[t]{0.3\textwidth}
\centering
% --- Plasmon decay ---
\begin{tikzpicture}[scale=0.5]
\begin{feynman}
\vertex (a) {\(\gamma_{\rm L,T}\)};
\vertex [right=of a] (b);
\vertex [above right=of b] (f1) {\(\chi\)};
\vertex [below right=of b] (f2) {\(\overline{\chi}\)};
\diagram* {
(a) -- [photon] (b) -- [fermion] (f1),
(f2) -- [fermion] (b),
};
\end{feynman}
\end{tikzpicture}

\small (Plasmon) {\bf{D}}ecay
\end{minipage}

\end{tabular}

\caption{Processes for millicharged particles pair production. }
\label{fig:feynmandiagrams}
\end{figure*}

The dispersion relation of photons is modified in the dense plasma,
allowing for their decay even in the context of the Standard Model~\cite{Adams:1963zzb,1965NCimA..40..502Z,Beaudet:1967zz,Dicus:1972yr,Braaten:1993jw,Haft:1993jt,Raffelt:1996wa,Ratkovic:2003td,Vitagliano:2017odj}. Although for most processes the in-medium corrections are negligible---since $\omega_{\rm pl}\ll T$, photons are ultra-relativistic and can be treated as massless---we will of course include them when dealing with plasmon decay; only for this reaction one needs to include the renormalized medium properties. The information on the medium-induced dispersion is contained in the self-energy, which is different for the longitudinal and the transverse states of the photon field. It is convenient to regard these as two separate species altogether; in the Lorentz gauge, for a photon moving along the direction $x$ and four-vector $K^\mu=(\omega,k,0,0)$, the longitudinal mode has a single polarization vector $e_L^\mu=(k,\omega,0,0)/\sqrt{\omega^2-k^2}$, while the transverse mode has two separate states $e_{T,1}^\mu=(0,0,1,0)$ and $e_{T,2}^\mu=(0,0,0,1)$.

For the longitudinal field, the self-energy is~\cite{Haft:1993jt,Raffelt:1996wa}
\begin{equation}
    \pi_L(\omega,k)=\frac{\omega^2-k^2}{\omega^2}\omega_{\rm pl}^2,
\end{equation}
where we have assumed $v_{\rm T}k\ll\omega_{\rm pl}$ where $v_{\rm T}$ is the typical thermal velocity of the electrons---plasmons with higher momentum are rapidly Landau-damped. Hence, the propagator of the longitudinal photon in Lorentz gauge is altered to 
\begin{equation}
    \frac{1}{\omega^2-k^2}\to \frac{\omega^2}{(\omega^2-k^2)(\omega^2-\omega_{\rm pl}^2)}.
\end{equation}
The pole $\omega=k$ is not physical---its polarization vector $e_L^\mu$ would vanish on contraction with any other physical current due to gauge invariance---so the dispersion relation of the longitudinal state is simply $\omega=\omega_{\rm pl}$ with a wavefunction renormalization $Z_L=\omega_{\rm pl}^2/(\omega_{\rm pl}^2-k^2)$. For the transverse state, we have even more simply $\pi_T=\omega_{\rm pl}^2$, so that there is no wavefunction renormalization $Z_T=1$ and the dispersion relation is that of a massive particle $\omega^2=k^2+\omega_{\rm pl}^2$.

Millicharged particles can be produced by different processes depending on the temperature and density of the plasma, akin to neutrinos produced in thermal interactions, for which different production processes can be comparable for certain plasma parameters~\cite{Haft:1993jt}, as is the case for the Sun~\cite{Vitagliano:2017odj}. Moreover, the MCP mass is unknown; therefore, some processes can be kinematically forbidden depending on the value of $m_\chi$. 
The rates for energy cooling are in principle due to the ABCD processes: (pair) \textbf{A}nnihilation, \textbf{B}remsstrahlung, \textbf{C}ompton, and (plasmon) \textbf{D}ecay (see Fig.~\ref{fig:feynmandiagrams}).\footnote{This is akin to the ABC processes for axion~\cite{Redondo:2013wwa} and the ABCD processes for neutrino~\cite{Vitagliano:2017odj,Vitagliano:2017ona} production in the Sun, except from the different meaning of the \textbf{A} reaction which stands there for ``atomic transitions''---the latter being negligible at the large temperatures considered in this work.} In practice, bremsstrahlung production is subdominant in the stellar environments relevant for this work. Millicharged particles are produced through a vector current, analogously to neutrinos, and as a result their production scales similarly to neutrinos in the same plasma conditions. In the late evolutionary stages of massive stars, with central temperatures $T\gtrsim 10\,\rm{keV}$ and densities $\rho \gtrsim 10^3\,\rm g \,cm^{-3}$, neutrino production via bremsstrahlung has been shown to be subdominant compared to Compton and plasmon processes (see, e.g., Fig.~1 in Ref.~\cite{Haft:1993jt}). Bremsstrahlung would become relevant only in environments with significantly higher densities at comparable or lower temperatures, such as degenerate stellar cores, which are not the focus of this study. We therefore neglect bremsstrahlung and restrict our analysis to the pair annihilation, Compton, and plasmon decay processes in the following.

\begin{widetext}
\section{Plasmon decay}
\label{sec:pldec}
For light MCPs, when $m_\chi < \omega_{\rm pl}/2$, the dominant production channel is the plasmon decay $\gamma_{L,T}\to\chi \overline{\chi}$ (the D process).
The energy-loss rate for this process can be written as
\begin{equation}
   \frac{d\mathcal{E}_{\gamma\to\chi\overline{\chi}}}{dVdt} =  \int \frac{d^3\bp_1}{(2\pi)^{3}2E_1} \frac{d^3\bp_2}{(2\pi)^{3}2E_2}\frac{d^3\bk}{(2\pi)^{3}2\omega} (2\pi)^4\,\delta^{(4)}(K-P_1-P_2)\,\big|\mathcal{M}_{\gamma\to \chi\overline{\chi}}\big|^2\,\omega\,f(\omega)Z(\omega(k))\,,
\end{equation}

where $P_i=(E_i,\,\bp_i)$ are the four momenta of the MCPs, $K=(\omega,\bk)$ is the four momentum of the plasmon, with $\omega$ and $k$ related through the plasmon dispersion relation reviewed in Sec.~\ref{sec:plasma}, and the factor $\omega$ accounts for the energy lost in the production of the pairs, equal to the entire energy of the decaying plasmon. In this expression, $f(\omega)$ is the Bose-Einstein distribution for the plasmons with energy $\omega$, $Z(\omega,k)$ is the wavefunction renormalization factor introduced in Sec.~\ref{sec:plasma}, and $|\mathcal{M}_{\gamma\to \chi\overline{\chi}}|^2$ is the squared matrix element for plasmon decay

\begin{equation}
    |\mathcal{M}_{\gamma\to \chi\overline{\chi}}|^2=16\pi q^2\alpha e_\mu e_\nu \left[P_1^\mu P_2^\nu+P_1^\nu P_2^\mu-(m_\chi^2+P_1\cdot P_2)g^{\mu\nu}\right],
\end{equation}
with $e_\mu$ the polarization vector of the plasmon.

The cooling rate for a given polarization state is now
\begin{equation}
    \frac{d\mathcal{E}_{\gamma\to\chi\overline{\chi}}}{dVdt}=16\pi q^2 \alpha\int \frac{d^3\bk}{(2\pi)^32\omega}\omega f(\omega)Z(\omega,k)e^\mu e^\nu T_{\mu\nu},
\end{equation}
where
\begin{equation}\label{eq:tensor}
    T^{\mu\nu}=\int\frac{d^3\bp_1}{(2\pi)^32E_1}\frac{d^3\bp_2}{(2\pi)^32E_2}(2\pi)^4\delta^{(4)}(K^\mu-P_1^\mu-P_2^\mu)\left[P_1^\mu P_2^\nu+P_1^\nu P_2^\mu-(m_\chi^2+P_1\cdot P_2)g^{\mu\nu}\right].
\end{equation}
We can easily verify by explicit computation that $T^{\mu\nu} K_\mu K_\nu=0$, which in fact descends also from the Ward identity. Hence, the tensor $T^{\mu\nu}$ must be written as
\begin{equation}
    T^{\mu\nu}=I\left[g^{\mu\nu}-\frac{K^\mu K^\nu}{K^2}\right].
\end{equation}
To obtain the value of $I$, we contract Eq.~\eqref{eq:tensor} with $g^{\mu\nu}$, so
\begin{equation}
    I=-\frac{2}{3}\int \frac{d^3\bp_1}{(2\pi)^32E_1}\frac{d^3\bp_2}{(2\pi)^32E_2}(2\pi)^4\delta^{(4)}(K^\mu-P_1^\mu-P_2^\mu)\left[2m_\chi^2+P_1\cdot P_2\right]\,,
\end{equation}
which can be rewritten as
\begin{equation}
    I=-\frac{2}{3}\left(m_\chi^2+\frac{K^2}{2}\right)\int \frac{d^3\bp_1}{(2\pi)^32E_1}\frac{d^3\bp_2}{(2\pi)^32E_2}(2\pi)^4\delta^{(4)}(K^\mu-P_1^\mu-P_2^\mu).
\end{equation}

The integral can now be performed explicitly, so that we find
\begin{equation}
    I=-\frac{2}{3}\left(m_\chi^2+\frac{K^2}{2}\right)\frac{\sqrt{K^2-4m_\chi^2}}{8\pi K}.
\end{equation}

Finally, by replacing the polarization vector for the longitudinal state, we obtain the cooling rate due to longitudinal plasmon decay
\begin{equation}
\frac{d\mathcal{E}_{\gamma_L\to\chi\overline{\chi}}}{dVdt} = \frac{q^2\,\alpha\,\omega_{\rm pl}^2}{6\,\pi^2}
\int_0^{\sqrt{{\omega_{\rm pl}^\prime}^2-4m_\chi^2}}\, dk k^2 \sqrt{1-\frac{4\,m_\chi^2}{\omega_{\rm pl}^2-k^2}}\frac{\omega_{\rm pl}^2 -k^2 + 2m_\chi^2}{(\omega_{\rm pl}^2-k^2)(e^{\omega_{\rm pl}/T}-1)}\,.
\end{equation}
For the transverse state, after summing over both polarizations, we find
\begin{equation}
\label{eq:QTpl}
    \frac{d\mathcal{E}_{\gamma_T\to\chi\overline{\chi}}}{dVdt} = \frac{q^2\,\alpha}{3\,\pi^2}\sqrt{1-\frac{4\,m_\chi^2}{\omega_{\rm pl}^2}}(\omega_{\rm pl}^2 + 2m_\chi^2)\,\int\,\frac{k^2 dk}{e^{\sqrt{\omega_{\rm pl}^2+k^2}/T}-1}\,.
\end{equation}
The result of the integral in Eq.~\eqref{eq:QTpl} is a function of only $\omega_{\rm pl}/T$. Therefore, we can express the emissivity as
\begin{equation}\label{eq:transverse_fit}
    \frac{d\mathcal{E}_{\gamma_{T}\to\chi\overline{\chi}}}{dVdt} = \frac{q^2\,\alpha\,T^3}{3\,\pi^2}\sqrt{1-\frac{4\,m_\chi^2}{\omega_{\rm pl}^2}}(\omega_{\rm pl}^2 + 2m_\chi^2)\,\Phi_T(\omega_{\rm pl}/T)\,,
\end{equation}
where
\begin{equation}
  \Phi_T(x) = \exp[0.889-0.346\,x^{1.314}]\,. 
\end{equation}
\end{widetext}

At this stage, we note that the obtained expressions are applicable provided the plasma is non-relativistic and non-degenerate.~There are no further conditions, and in particular they hold both for $T\gg\omega_{\rm pl}$---the regime of interest for hot, pre-supernova stellar cores---and $T\lesssim \omega_{\rm pl}$---a condition interesting for the cores of red giants and the early stages of white dwarfs.~It is instructive to determine which of the two polarization states---longitudinal or transverse---dominates in each regime.

For the case $T\ll \omega_{\rm pl}$, these expressions simplify significantly in the low-mass MCP regime; for the longitudinal state we find
\begin{equation}
    \frac{d\mathcal{E}_{\gamma_L\to \chi\overline{\chi}}}{dVdt}=\frac{q^2\alpha \omega_{\rm pl}^5}{18\pi^2}e^{-\omega_{\rm pl}/T},
\end{equation}
while for the transverse states 
\begin{equation}
    \frac{d\mathcal{E}_{\gamma_T\to \chi\overline{\chi}}}{dVdt}=\frac{q^2\alpha \omega_{\rm pl}^2}{12}e^{-\omega_{\rm pl}/T}\left(\frac{2\omega_{\rm pl}T}{\pi}\right)^{3/2}.
\end{equation}
Hence, for $T\ll \omega_{\rm pl}$, the contribution from longitudinal plasmons dominates the emission by a factor $\sim(\omega_{\rm pl}/T)^{3/2}$.

In the opposite limit $\omega_{\rm pl}\ll T$, which is more interesting for us, the massless regime is equally simple and gives
\begin{equation}
\begin{aligned}
    \frac{d\mathcal{E}_{\gamma_L\to \chi\overline{\chi}}}{dVdt}&=\frac{q^2 \alpha\omega_{\rm pl}^4 T}{18\pi^2},\,  \\
    \frac{d\mathcal{E}_{\gamma_T\to \chi\overline{\chi}}}{dVdt}&=\frac{2q^2\alpha \zeta(3)\omega_{\rm pl}^2 T^3}{3\pi^2},
\end{aligned}
\end{equation}
showing explicitly that the transverse contribution dominates by a factor $\sim (T/\omega_{\rm pl})^2$. With this motivation, we will include only the transverse contribution through the fit in Eq.~\eqref{eq:transverse_fit}.

\section{Compton production}
\label{sec:compt}

At higher masses, when $m_\chi > \omega_{\rm pl}/2$, plasmon decay is no longer kinematically allowed. In this context, the relevant MCP production process is the Compton-like scattering $e^-+\gamma\to e^-+\overline{\chi}+\chi$ (the D process), a $2\to 3$ process which may be relevant over a wide range of temperatures, from $T\ll m_e$ up to $T \gg m_e$.  Compton scattering off nucleons is generically subdominant because of their larger inertia.

Before proceeding with the actual calculations, let us briefly discuss our assumptions about the energy scales involved in the problem. In the following, we assume $m_\chi \gg \omega_{\rm pl}$. As discussed in Sec.~\ref{sec:pldec}, in the opposite regime, for $m_\chi\lesssim\omega_{\rm pl}$, MCPs can be produced by the on-shell decay of plasmons. Instead, in the regime of interest to us, Compton production $e^-+\gamma\to e^++\overline{\chi}+\chi$ via off-shell photons decaying into a pair of MCPs is the dominant reaction. In principle, one could write expressions that are valid throughout the two regimes; this was attempted by Refs.~\cite{Vogel:2013raa, Fung:2023euv}. The general idea is that the photon may be represented by a propagator which, in the limit of an infinitely small imaginary part, turns into a delta function enforcing the on-shell condition, leading to plasmon decay, while in the off-shell regime describes Compton scattering. However, Refs.~\cite{Vogel:2013raa,Fung:2023euv} describe the imaginary part of the self-energy for the photon as if it was on-shell. This is certainly a good approximation for the plasmon decay regime, but in that case the imaginary part is irrelevant, as it is only a small width for the delta function. In the Compton scattering regime, this approximation is unjustified, as the intermediate photon is of course off-shell---an on-shell plasmon in this regime cannot decay into a $\overline{\chi}\chi$ pair. Nevertheless, we will find that for the transverse states the imaginary part of the self-energy is indeed the same also for off-shell photons, so that we find the same results as Ref.~\cite{Fung:2023euv} in the non-relativistic plasma regime. On the other hand, for the longitudinal component we find a considerably different result, as we argue below. Finally, no previous work has attempted to bridge the Compton emissivity into the relativistic QED plasma regime.

Motivated by this, we will assess here the production in the Compton regime independently. 
We will rely on the approximation that $m_\chi\gg \omega_{\rm pl}$. Furthermore, we will only consider non-degenerate plasmas, which for massive stars is an excellent assumption except only for the very final stages of their evolution. Under these approximations, the contribution from incoming longitudinal plasmons can be neglected; in fact, for a non-relativistic plasma, the electron is recoilless, so that the plasmon should have an energy $E_\gamma\gtrsim m_\chi\gg\omega_{\rm pl}$, which is impossible for an on-shell plasmon. For a relativistic plasma, with $T\sim m_e$, the thermal electron velocity approaches the speed of light, so that most plasmons are ultra-relativistic with $\omega\simeq k$. Such longitudinal plasmons decouple from the medium due to gauge invariance; by the same argument, longitudinal plasmons decouple from incoherent processes also in supernova cores~\cite{Fiorillo:2024upk}. Overall, in all regimes we are allowed to neglect longitudinal plasmons as on-shell particles. The corrections to the photon propagator due to the medium can then be discarded altogether; virtual photons are off-shell by an energy amount much larger than $\omega_{\rm pl}$. The neglect of longitudinal plasmons as incoming degrees of freedom already makes it clear that we will find results different from those of Refs.~\cite{Vogel:2013raa,Fung:2023euv}: these works use, for the intermediate longitudinal plasmon decaying into millicharged particles, a self-energy from Ref.~\cite{Redondo:2008aa} that accounts only for $\gamma_Le^-\to \gamma_Le^-$. (This self-energy was anyway also incorrect, as pointed out by several authors later~\cite{An:2013yfc,Redondo:2013lna}.)

For a non-relativistic plasma, there is an additional question to consider, namely the possibility of screening. In fact, the spatial scale of screening is described by the Debye scale $k_{\rm D}=\omega_{\rm pl}/v_{\rm T}$, where $v_{\rm T}$ is the thermal velocity of the electrons. Since for a non-relativistic plasma $v_{\rm T}\sim \sqrt{T/m_e}$, for $T\ll m_e$ this scale can be much more relevant than the plasma frequency. However, it is easy to see that the Debye scale makes its appearance only for fields which vary much faster in space than in time, i.e. with $|\bk|$ (the modulus of the photon wavevector) much larger than $\omega$ (the frequency of the photon). This is the only regime in which powers of $v^{-1}$, where $v$ is the electron velocity, may appear. For example, in Eqs.~(6.37-6.38) of Ref.~\cite{Raffelt:1996wa}, it is immediately clear that the powers of $v^{-1}$ can only appear when an expansion in powers of $\omega/k$ is performed. However, the wavevector of the virtual photon decaying into $\overline{\chi}\chi$ is always strongly time-like, with $\omega \gg |\bk|$. Therefore, the screening scale simply does not appear.

In order to proceed with the computation, let us denote by $\bp_1$ and $\bp_2$ the initial and final momentum of the electron, and the corresponding energies are $E_1$ and $E_2$. The photon momentum is $\bq$, while the momenta and energies of the millicharged particles are $\bk_1$, $\bk_2$ and $\varepsilon_1$, $\varepsilon_2$.  Finally, we denote by $P^\mu=k_1^\mu+k_2^\mu$ the total four-momentum of the MCPs, and by $\xi^\mu=(k_1^\mu-k_2^\mu)/2$ their relative four-momentum.
It is convenient to consider, rather than the volumetric cooling rate, more generally the amount of four-momentum lost by the plasma,  written as
\begin{equation}\label{eq:general_cooling_rate}
    \frac{dP^\sigma}{dVdt}=\int \frac{d^3\bp_1}{(2\pi)^3 2E_1}\int \frac{d^3\bq}{(2\pi)^3 2|\bq|}f_{e^-}(E_1) f_\gamma(|\bq|)\mathcal{F}^\sigma,
\end{equation}
where $f_{e^-}$ and $f_\gamma$ are the electron and photon distribution functions and

\begin{widetext}
\begin{align}\label{eq:def_F_sigma}
    \mathcal{F}^\sigma=\int \frac{d^3\bp_2}{(2\pi)^3 2E_2}\frac{d^3\bk_1}{(2\pi)^32\varepsilon_1}\frac{d^3\bk_2}{(2\pi)^32\varepsilon_2}(2\pi)^4\delta(P^\mu+p_2^\mu-p_1^\mu-q^\mu) P^\sigma|\mathcal{M}|^2.
\end{align}

We do not include any spin or polarization factor, so the matrix element is assumed to be summed over spins and polarizations.
The matrix element can be factorized into a part relating only to the MCPs, and a part relating to the electron-photon system. Specifically we have
\begin{equation}
    |\mathcal{M}|^2=\frac{(4\pi\alpha)^3q^2}{P^4}\mathcal{P}_{\alpha\beta}\mathcal{T}_{\mu\nu}\mathcal{L}^{\mu\nu,\alpha\beta}.
\end{equation}

Here,
\begin{equation}
    \mathcal{P}_{\alpha\beta}=-g_{\alpha\beta}
\end{equation}
is the polarization tensor of the incoming photon summed over all polarizations (in the massless limit considered here, the longitudinal state decouples, so these are purely transverse), while
\begin{equation}
    \mathcal{T}^{\mu\nu}=4[k_1^\mu k_2^\nu+k_1^\nu k_2^\mu-(k_1\cdot k_2+m_\chi^2)g^{\mu\nu}]=4\left[\frac{P^\mu P^\nu-P^2 g^{\mu\nu}}{2}-2\xi^\mu \xi^\nu\right]
\end{equation}
is the spin-summed trace over the millicharged spinors. Finally
\begin{equation}
\mathcal{L}^{\mu\nu,\alpha\beta}=\mathcal{L}_{(1)}^{\mu\nu,\alpha\beta}+\mathcal{L}_{(2)}^{\mu\nu,\alpha\beta}+\mathcal{L}_{(3)}^{\mu\nu,\alpha\beta}
\end{equation}
is the lepton trace, separated in its t-channel, u-channel, and interference contribution

\begin{eqnarray}
    \mathcal{L}^{\mu\nu,\alpha\beta}_{(1)}&=&\frac{\mathrm{Tr}[\gamma^\mu (\slashed{p}_1+\slashed{q}+m_e)\gamma^\alpha(\slashed{p}_1+m_e)\gamma^\beta (\slashed{p}_1+\slashed{q}+m_e) \gamma^\nu (\slashed{p}_2+m_e)]}{[(p_1+q)^2-m_e^2]^2},\\ \nonumber
\mathcal{L}^{\mu\nu,\alpha\beta}_{(2)}&=&\frac{\mathrm{Tr}[\gamma^\alpha (\slashed{p}_2-\slashed{q}+m_e)\gamma^\mu(\slashed{p}_1+m_e)\gamma^\nu (\slashed{p}_2-\slashed{q}+m_e) \gamma^\beta (\slashed{p}_2+m_e)]}{[(p_2-q)^2-m_e^2]^2},\\ \nonumber
\mathcal{L}^{\mu\nu,\alpha\beta}_{(3)}&=&\frac{2\mathrm{Tr}[\gamma^\mu (\slashed{p}_1+\slashed{q}+m_e)\gamma^\alpha(\slashed{p}_1+m_e)\gamma^\nu (\slashed{p}_2-\slashed{q}+m_e) \gamma^\beta (\slashed{p}_2+m_e)]}{[(p_1+q)^2-m_e^2] [(p_2-q)^2-m_e^2]}.
\end{eqnarray}

In Eq.~\eqref{eq:def_F_sigma}, we can now use the relativistic invariance of the integrand to explicitly integrate out the variable $\xi^\mu$, leaving only an integral over the total MCP momentum $P^\mu$. To do this, we use the identity

\begin{equation}
    \int \frac{d^3\bk_1}{(2\pi)^3 2\varepsilon_1}\frac{d^3\bk_2}{(2\pi)^3 2\varepsilon_2}=\int \frac{d^4P}{(2\pi)^3}\frac{d^4\xi}{(2\pi)^3}\frac{1}{2}\delta\left(\frac{P^2}{4}+\xi^2-m_\chi^2\right)\delta(P\cdot \xi);
\end{equation}
the integral must be performed only over the regions such that $k_1^0>0$ and $k_2^0>0$. Finally, from the two identities 
\begin{align}
    \int \frac{d^3\bk_1}{(2\pi)^3 2\varepsilon_1} \frac{d^3\bk_2}{(2\pi)^3 2\varepsilon_2}&=\int \frac{d^4P}{64\pi^5 P}\sqrt{\frac{P^2}{4}-m_\chi^2}\,,\\
    \int \frac{d^3\bk_1}{(2\pi)^3 2\varepsilon_1} \frac{d^3\bk_2}{(2\pi)^3 2\varepsilon_2} \xi^\mu \xi^\nu&=\int \frac{d^4P}{64\pi^5P }\sqrt{\frac{P^2}{4}-m_\chi^2}\frac{\frac{P^2}{4}-m_\chi^2}{3P^2}(P^\mu P^\nu-P^2 g^{\mu\nu})\,,
\end{align}
we are led to the expression
\begin{align}
\label{eq:general_cooling_compton}
    \mathcal{F}^\sigma=\int \frac{d^3\bp_2}{(2\pi)^3 2E_2} \frac{d^4 P}{64 \pi^5}(2\pi)^4\delta(P^\mu+p_2^\mu-p_1^\mu-q^\mu)\frac{(4\pi\alpha)^3 q^2 \mathcal{P}_{\alpha\beta}\mathcal{L}^{\mu\nu,\alpha\beta}}{P^4}P^\sigma \frac{\sqrt{\frac{P^2}{4}-m_\chi^2}}{P}2(P_\mu P_\nu-P^2 g_{\mu\nu})\frac{2+\frac{4m_\chi^2}{P^2}}{3}.
\end{align}

The integral over $P^0$ runs only from $0$ to $+\infty$, and furthermore we must have $P^2>4m_\chi^2$ (physically, the pairs have a minimum energy when they are produced at rest).
In this way, we have managed to reduce the cooling rate to a standard emission of a photon with four-momentum $P^\mu$, but with an arbitrary dispersion relation, so that $P^2$ does not vanish.

Overall, the rate of energy-momentum loss from the medium can be written as

\begin{align}\label{eq:rate_energy_momentum_loss}
    \frac{dP^\sigma}{dVdt}=\int \frac{d^4P}{(2\pi)^5 }\left(\frac{P_\mu P_\nu}{P^2}-g_{\mu\nu}\right)\frac{2}{3P^2}\left(1+\frac{2m_\chi^2}{P^2}\right)\frac{\sqrt{\frac{P^2}{4}-m_\chi^2}}{P}P^\sigma \Gamma^{\mu\nu}(P),
\end{align}
  where
\begin{equation}
    \Gamma^{\mu\nu}=\int \frac{d^3\bp_1}{(2\pi)^32E_1}\frac{d^3\bp_2}{(2\pi)^3 2E_2}\frac{d^3\bq}{(2\pi)^32|\bq|} f_{e^-}(E_1) f_\gamma(|\bq|)(2\pi)^4\delta(P^\mu+p_2^\mu-p_1^\mu -q^\mu) (4\pi\alpha)^3 g^2 \mathcal{P}_{\alpha\beta}\mathcal{L}^{\mu\nu,\alpha\beta}.
\end{equation}

In this form, Eq.~\eqref{eq:rate_energy_momentum_loss} is analogous to the one used in Ref.~\cite{Vogel:2013raa}; it accounts for production from transverse photons with four-momentum $P^\mu$, generally off-shell so that $P^2\neq 0$, with a purely transverse polarization tensor proportional to $P_\mu P_\nu-P^2 g_{\mu\nu}$, and an interaction rate for the photons determined by the function $\Gamma^{\mu\nu}$. In fact, if we introduce the average polarization tensor
\begin{equation}
    \langle e_\mu e_\nu\rangle=\frac{1}{2}\left[\frac{P_\mu P_\nu}{P^2}-g_{\mu\nu}\right],
\end{equation}
with the normalization chosen to correspond to a single polarization state in vacuum, we can define the scattering rate for a single polarization state as
\begin{equation}
    \gamma=\frac{1}{2\omega}\frac{\Gamma^{\mu\nu}\langle e_\mu e_\nu \rangle}{4\pi \alpha q^2},
\end{equation}
where the charge of the millicharged pair has been removed in order to isolate the scattering rate of the virtual photon and the factor $(2\omega)^{-1}$ corresponds to the normalization of the photon wavefunction. If we now write $P^0=\omega$, $|\bP|=k$ (to use the same notation as Refs.~\cite{Vogel:2013raa,Fung:2023euv}) we finally obtain

\begin{equation}
    \frac{d\mathcal{E}_{\gamma e^-\to e^- \overline{\chi}{\chi}}}{dVdt}=\frac{\alpha q^2}{3\pi^3}\int dk d\omega \frac{k^2 \omega(\omega^2-k^2+2m_\chi^2)\sqrt{\omega^2-k^2-4m_\chi^2}}{\sqrt{\omega^2-k^2}}\frac{2\gamma \omega}{(\omega^2-k^2)^2}.
\end{equation}
\end{widetext}

This coincides with the transverse contribution obtained by thermal field theory in Ref.~\cite{Fung:2023euv}, provided that $\gamma=\omega \mathrm{Im}\left[\Pi_T(\omega,k)\right]/f_\gamma(\omega)$ in their notation.

To move forward, let us finally introduce the quantity
\begin{equation}
    \Gamma(P)=\Gamma^{\mu\nu}\left(\frac{P_{\mu}P_\nu}{P^2}-g_{\mu\nu}\right)=16\pi \alpha q^2 \omega \gamma.
\end{equation}
Notice that this function is not a relativistic invariant, since it depends on the distribution functions $f_{e^-}(E_1)$ and $f_{\gamma}(|\bq|)$, which single out a laboratory reference frame in which the plasma is at rest. Therefore, we must evaluate $\Gamma(P)$ in the correct reference frame comoving with the plasma. We will also define $\gamma_T=\Gamma/2P^0$, which corresponds to the absorption rate of a virtual photon with an off-shell four-momentum $P^\mu$.

Let us separately discuss the non-relativistic (NR), where $T,m_\chi\ll m_e$, and the ultra-relativistic  (UR) limit, where $T\gg m_e$.

\subsection{Non-relativistic limit}

Let us define
\begin{equation}
    \mathcal{L}\equiv\mathcal{L}^{\mu\nu,\alpha\beta}\mathcal{P}_{\alpha\beta}(\frac{P_\mu P_\nu}{P^2}-g_{\mu\nu})\,,
\end{equation}
which is a relativistic invariant. In the function $\Gamma$, we can now perform the relevant integrals over the delta function. In the NR limit, we have $\bp_2=\bp_1+\bq-\bP$, while the energy-conservation delta simply gives us $|\bq|=P^0$. Therefore, we find
\begin{equation}
    \Gamma=\frac{(4\pi\alpha)^3 q^2 P^0}{(2\pi)^5 8m_e^2 }\int d^3\bp_1 d\Omega_\bq f_{e^-}(E_1) f_\gamma(|\bq|) \mathcal{L},
\end{equation}
where $d\Omega_\bq$ is the solid angle of the vector $\bq$. To determine $\mathcal{L}$, it is convenient to proceed in the center-of-mass (COM) frame. This requires us to relate the momenta of the particles in the laboratory frame (where by definition $P^\mu=(P^0,P^1,0,0)$, since we may choose the axes aligned with the direction of $\bP$) and in the COM frame.
We may therefore denote by $|\tilde{\bq}|$ the modulus of the photon momentum, $|\tilde{\bP}|$ the modulus of the virtual photon momentum, and by $\tilde{x}$ the cosine of the angle between the two vectors, in the COM frame. Then, to lowest order in the NR expansion we can compute 
\begin{equation}
    \mathcal{L}=\frac{8\left[2\tq^2-\tP^2(1-\tx^2)\right]}{\tq^2}.
\end{equation}
To verify the correctness of this expression, we notice that for a massless photon we must have $\tP=\tq$, for which $\mathcal{L}=8(1+\tx^2)$, corresponding to the standard angular distribution for the photons produced via Thomson scattering. 

We must now relate $\tq$, $\tP$, and $\tx$, with the corresponding quantities in the laboratory frame. In the NR limit, we easily see that $\tq=|\bq|$ from the invariance of the Mandelstam parameter $s$ and $\tP=\sqrt{|\bq|^2-P^2}$ from the invariance of $P^2$. Finally, from the invariance of $P\cdot q$, we find that
\begin{equation}
    \tP \tx=-P^0+P^1 z+|\bq|,
\end{equation}
where $z$ is the cosine of the angle between $\bq$ and $\bP$ in the laboratory frame. After replacing, the angular integrations are now trivial. In order to match with the notation of Refs.~\cite{Vogel:2013raa,Fung:2023euv}, we will now write $\omega=P^0$, $k=P^1$, and we are finally led to the expression
\begin{equation}
    \Gamma=\frac{64\pi^2\alpha^3 q^2 n_e \omega \left(1-\frac{k^2}{3\omega^2}\right) f_\gamma(\omega)}{m_e^2}.
\end{equation}
It is instructive to also separate out the contribution due to the transverse and the longitudinal polarization states of the virtual photon. To do this, we rewrite the tensor
\begin{equation}
    \frac{P^\mu P^\nu}{P^2}-g^{\mu\nu}=\sum_{i}e_i^\mu e_i^{*,\nu}\,,
\end{equation}
for the three polarization vectors of the field. By explicitly writing the transverse and the longitudinal vectors, we may separate $\Gamma=\Gamma_T+\Gamma_L$, with the transverse rate being
\begin{equation}
    \Gamma_T=\frac{128\pi^2\alpha^3 q^2 n_e \omega  f_\gamma(\omega)}{3m_e^2},
\end{equation}
while the longitudinal one is
\begin{equation}\label{eq:Im_PiL_plasmon}
    \Gamma_L=\frac{64\pi^2\alpha^3 q^2 n_e \omega \left(1-\frac{k^2}{\omega^2}\right) f_\gamma(\omega)}{3m_e^2}.
\end{equation}

Hence, the cooling rate can finally be written in the form
\begin{widetext}

\begin{equation}
    \frac{d\mathcal{E}_{\gamma e^-\to e^- \overline{\chi}{\chi}}}{dVdt}=\int d^4 P f_\gamma(\omega)\frac{2\alpha^3 g^2 n_e \sqrt{\omega^2-k^2-4m_\chi^2}(\omega^2-k^2+2m_\chi^2)(3\omega^2-k^2)}{9\pi^3 m_e^2 (\omega^2-k^2)^{5/2}}.
\end{equation}

After expanding the integral over $P$ we obtain

\begin{equation}\label{eq:cooling_rate}
    \frac{d\mathcal{E}_{\gamma e^-\to e^- \overline{\chi}{\chi}}}{dVdt}=\int dk d\omega f_\gamma(\omega)\frac{8\alpha^3 g^2 n_e k^2 \sqrt{\omega^2-k^2-4m_\chi^2}(\omega^2-k^2+2m_\chi^2)(3\omega^2-k^2)}{9\pi^2 m_e^2 (\omega^2-k^2)^{5/2}}.
\end{equation}

 If we had used $\Gamma_T$ in place of $\Gamma$, we would have recovered exactly the transverse contribution from Ref.~\cite{Fung:2023euv} in the limit of off-shell photons (i.e. with $|\omega^2-k^2-\mathrm{Re}(\Pi_T)|\gg|\mathrm{Im}(\Pi_T)|$ in Ref.~\cite{Fung:2023euv}). This shows that the scattering rate for off-shell photons is ultimately the same as for on-shell photons, a fact that is not trivial \textit{a priori} and was taken as an implicit assumption in Refs.~\cite{Vogel:2013raa,Fung:2023euv};~our calculation ultimately proves it. However, the contribution coming from longitudinal photons is entirely different from the one considered in Ref.~\cite{Fung:2023euv}. This is ultimately to be tracked to the different expressions for the self-energy of the longitudinal plasmons, namely our Eq.~\eqref{eq:Im_PiL_plasmon} and Eq.~(9) in Ref.~\cite{Fung:2023euv}. In the Lorentz gauge, the corresponding expression should vanish for $\omega=k$, a consequence of gauge invariance since in this limit the longitudinal polarization vector aligns with $P^\mu$. (We stress that the physical interaction rate of a longitudinal plasmon with $\omega=k$ may still be non-vanishing, since it should be multiplied by a wavefunction renormalization $\mathcal{Z}_L$ which diverges for $\omega=k$; but the self-energy itself must necessarily vanish in this limit.) Hence, the self-energy adopted in Ref.~\cite{Fung:2023euv} cannot be correct; this was extracted from Ref.~\cite{Redondo:2008aa}, which was later amended by several works~\cite{An:2013yfc,Redondo:2013lna}. In any case, the effect on the constraints deduced in Ref.~\cite{Fung:2023euv} is presumably mild, due to the dominance of the contribution from transverse plasmons.

By rescaling all variables by a factor $T$, we pass to dimensionless variables and obtain
\begin{equation}
    \frac{d\mathcal{E}_{\gamma e^-\to e^- \overline{\chi}{\chi}}}{dVdt}=\frac{8\alpha^3 q^2 n_e T^4}{9\pi^2 m_e^2}\Lambda(m_\chi/T)
    \equiv q^2 n_e f_{\rm NR}(m_\chi,T)\,,
\label{eq:Compt_NR}
\end{equation}
with
\begin{equation}
    f_{\rm NR}(m_\chi,\,T) = \frac{8\,\alpha^3\,T^4}{9\,\pi^2\,m_e^2}\,\Lambda(m_\chi/T)\,. 
\label{eq:fNR}
\end{equation}
and
\begin{equation}
    \Lambda(x)=\int\frac{dkd\omega}{e^\omega-1}\frac{k^2\sqrt{\omega^2-k^2-4x^2}(\omega^2-k^2+2x^2)(3\omega^2-k^2)}{(\omega^2-k^2)^{5/2}}.
\label{eq:NRfunc}
\end{equation}
An excellent fit for $\Lambda(x)$, within $\sim 10\%$ accuracy in the mass range $10^{-1}\lesssim x\lesssim 30$, is given by
\begin{equation}
    \Lambda(x) = \exp[3.18 - 1.53 x^{1.06}-0.2\,\ln(x)]\,,
\end{equation}
while at lower masses we use the asymptotic expression (with a precision better than $1\%$)
\begin{equation}
    \Lambda(x) = 2.87-12.98\ln(x)\, \quad \quad x\lesssim 10^{-1}.
\end{equation}
The temperature dependence of the Compton emissivity in the NR limit (for $T<m_e$ at fixed $\rho Y_e=10^{5}\,\rm g \,cm^{-3}$) is shown by the dashed lines in the left panel of Fig.~\ref{fig:EpsComp} for three different MCP masses. The right panel of Fig.~\ref{fig:EpsComp} displays the relative error of the functional fit to $\Lambda$ (blue line), defined as $1-{\rm fit}/\Lambda$, which reproduces the exact result to better than $10\%$ throughout the region of interest.

\subsection{Ultra-relativistic limit}

To compute the emissivity in the UR limit, let us express Eq.~\eqref{eq:general_cooling_compton}, which is a Lorentz four-vector, in the COM frame. We denote all quantities in this frame by a tilde. Here, the momenta of the incoming particles are $\tilde{\bp}_1=(-\tilde{q},0,0)$ and $\tilde{\bq}=(\tilde{q},0,0)$. From symmetry, it is apparent that the three-momentum loss will be directed along $\tilde{\bq}$, implying that the only non-vanishing components of $\tilde{\mathcal{F}}^\sigma$ in the COM frame are $\tilde{\mathcal{F}}^0$ and $\tilde{\mathcal{F}}^1$. The transformation that allows us to obtain the cooling rate in the laboratory frame is
\begin{equation} \mathcal{F}^0=\frac{(p_1+q)\tilde{\mathcal{F}}^0+\sqrt{p_1^2+q^2+2\bp_1\cdot\bq}\tilde{\mathcal{F}}^1}{\sqrt{2(p_1 q-\bp_1\cdot\bq)}}.
\end{equation}
In general, $\tilde{\mathcal{F}}^\sigma$ can be written as

\begin{equation}
    \tilde{\mathcal{F}}^\sigma=\int \frac{d^3\tilde{\bp}_2}{(2\pi)^3 2\ttp_2}\frac{d^4\ttP}{(2\pi)^5}(2\pi)^4\delta(\ttP^\mu+\ttp^\mu_2-\ttp^\mu_1-\ttq^\mu)\frac{2(4\pi\alpha)^3q^2}{3\ttP^2}\left(1+\frac{2m_\chi^2}{\ttP^2}\right)\frac{\sqrt{\frac{\ttP^2}{4}-m_\chi^2}}{\ttP}\ttP^\sigma \mathcal{L}.
\label{eq:tildef}
\end{equation}
%\end{widetext}
By explicit Feynman diagram computation, we find in the COM frame
\begin{equation}
    \mathcal{L}=\frac{4\left[4\ttq^2-4k\ttq(1+y)+k^2(5+2y+y^2)\right]}{k\ttq\left(1-y+\frac{m_e^2}{2k\ttq}\left(\frac{\ttq}{k}-1\right)\right)}\,,
\end{equation}
where we have used $\tP^\mu=(\omega,-k y,-k \sqrt{1-y^2},0)$, so that $\bp_2=(k y, k \sqrt{1-y^2},0)$. In the equation above, we have assumed the electron to be massless, except in the denominator where we had to include the mass correction to regularize the collinear divergence at $y=1$.\,\footnote{The physical origin of such a  divergence is that in the limit of massless electrons the intermediate particle can absorb the external photon while staying on-shell.} 

In order to explicitly compute the integral in Eq.~\eqref{eq:tildef}, let us first impose the energy conservation $\omega=2\ttq-k$, finding 

\begin{equation}
 \tilde{\mathcal{F}}^\sigma=\frac{8\alpha^3q^2}{3}
    \int_{0}^{\ttq-\frac{m_\chi^2}{\ttq}} k dk \int_{-1}^{+1} \frac{dy}{\ttP^2}\left(1+\frac{2m_\chi^2}{\ttP^2}\right)\frac{\sqrt{\frac{\ttP^2}{4}-m_\chi^2}}{\ttP}\ttP^\sigma \mathcal{L},
\end{equation}
where $\ttP^2=(2\ttq-k)^2-k^2$. Since the integral over $y$ is largely dominated by $y\simeq 1$,we can perform it with logarithmic precision to obtain
\begin{equation}
    \int_{-1}^{1}dy \mathcal{L}\simeq 16\left[\frac{\ttq}{k}+\frac{2k}{\ttq}-2\right]\log\left[\frac{4k\ttq}{m_e^2\left(\frac{\ttq}{k}-1\right)}\right].
\end{equation}

%\begin{widetext}
Finally, by performing a rescaling $k\to \ttq k$, the integrals become purely a function of $m_\chi/\ttq$ and $m_e/\tq$, which may be written in the form
\begin{equation}
    \tilde{\mathcal{F}}^\sigma=\frac{8\alpha^3 q^2 \ttq}{3}\left[\Phi^\sigma\left(\frac{m_\chi}{\ttq}\right)+\Psi^\sigma\left(\frac{m_\chi}{\ttq}\right)\log\left[\frac{\ttq^2}{m_e^2}\right]\right],
\end{equation}
with
\begin{eqnarray}
    \Phi^\sigma(m_\chi/\ttq)&=&\int_0^{1-\frac{m_\chi^2}{\ttq^2}}\frac{k dk}{\ttP^2}\left(1+\frac{2m_\chi^2}{\ttP^2}\right)\sqrt{\frac{1}{4}-\frac{m_\chi^2}{\ttP^2}}\frac{\ttP^\sigma}{\ttq}16\left[\frac{1}{k}+2k-2\right]\log\left[\frac{4k^2}{1-k}\right]\,,\\
    \Psi^\sigma(m_\chi/\ttq)&=&\int_0^{1-\frac{m_\chi^2}{\ttq^2}}\frac{k dk}{\ttP^2}\left(1+\frac{2m_\chi^2}{\ttP^2}\right)\sqrt{\frac{1}{4}-\frac{m_\chi^2}{\ttP^2}}\frac{\ttP^\sigma}{\ttq}16\left[\frac{1}{k}+2k-2\right].
\end{eqnarray}

The $\Psi^\sigma$ functions are in fact analytical and given by
\begin{eqnarray}
    \Psi^0(x)&=&\frac{-4+6x^2-3x^4+x^6-\sqrt{1-x^2}(-4+x^6)\,\log(\frac{\sqrt{1-x^2}+1}{x})}{\sqrt{1-x^2}}\,, \\
    \Psi^1(x)&=&\frac{20-40x^2+17x^4+3x^6-3\sqrt{1-x^2}(4-6x^4+x^6)\,\log(\frac{\sqrt{1-x^2}+1}{x})}{3\sqrt{1-x^2}} \,.
\end{eqnarray}
%\end{widetext}
Notice that the function $\Psi^1(x)$ is negative-definite. Indeed, since $y\simeq 1$ is strongly favored, the virtual photon transports momentum in the opposite direction than the incoming photon.
%%%%%
\begin{figure*}
    \centering
    \includegraphics[width=0.49\textwidth]{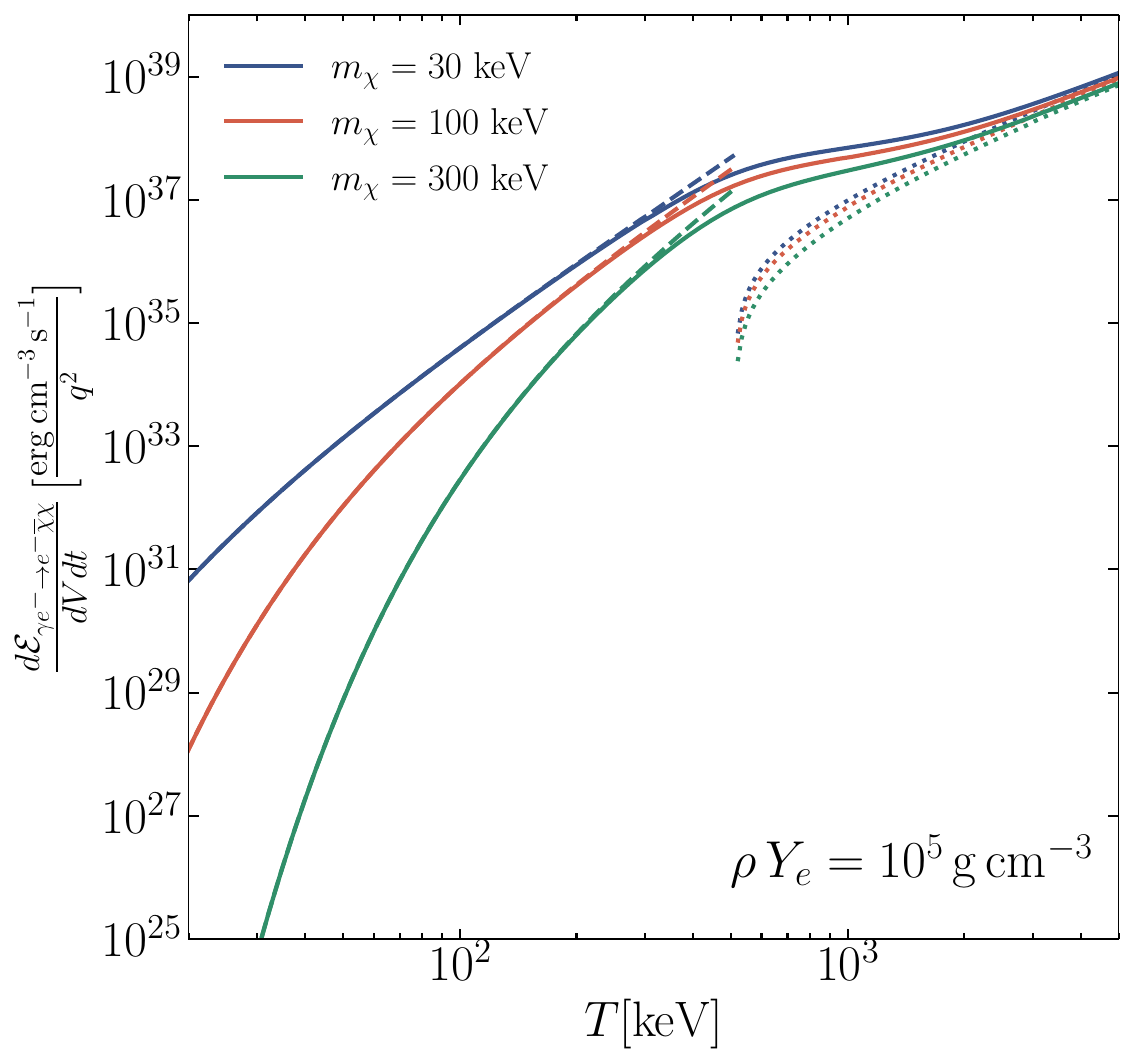}
    \includegraphics[width=0.49\textwidth]{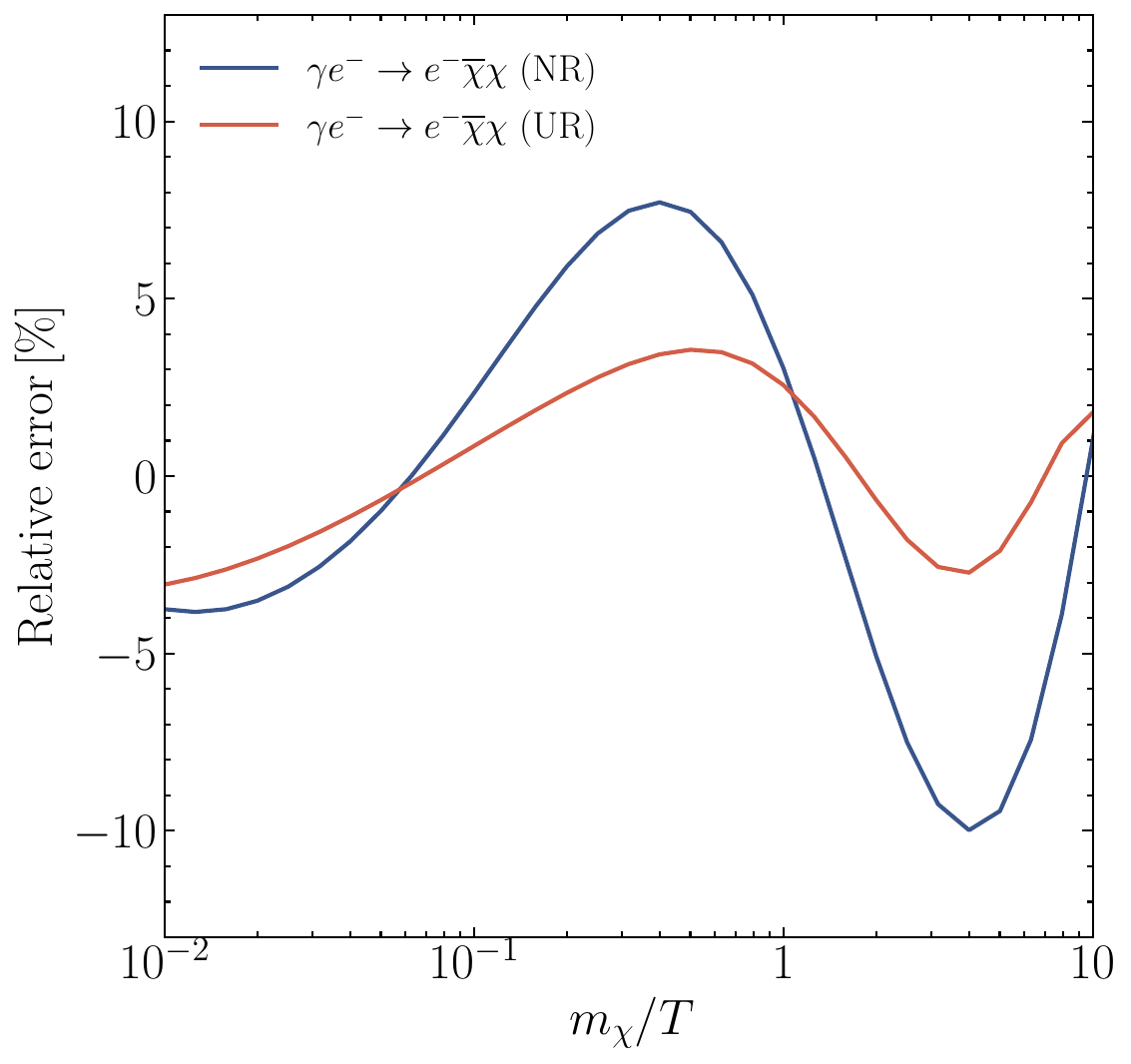}
    \caption{Left panel: MCP emissivity via Compton as a function of the temperature $T$ for different values of MCP masses at $\rho Y_e = 10^{5}\,\rm g \,cm^{-3}$. The dashed lines represent the non-relativistic limit given by Eq.~\eqref{eq:Compt_NR} while the ultra-relativistic limit given by Eq.~\eqref{eq:Compt_UR} is shown by the dotted lines. The solid lines show the emissivity given by Eq.~\eqref{eq:Compt_int}, obtained by interpolating the two limits. Right panel: Relative error in the fitting functions $\Lambda$ for the non-relativistic limit in Eq.~\eqref{eq:NRfunc} (blue curve) and $\Phi$ for the relativistic limit given by Eq.~\eqref{eq:URfunc} (red curve).}
    \label{fig:EpsComp}
\end{figure*}
%%%%%%
Moreover, the functions vanish for ${m_\chi/\ttq>1}$. 

Finally, in the logarithmic approximation $m_e\ll \ttq$, the logarithmic terms dominate, and therefore we may write more simply
\begin{equation}
    \tilde{\mathcal{F}}^\sigma=\frac{8\alpha^3 q^2 \ttq}{3}\Psi^\sigma\left(\frac{m_\chi}{\ttq}\right)\log\left[\frac{m_e^2}{\ttq^2}\right].
\end{equation}
Therefore, the cooling rate is
%\begin{widetext}
\begin{equation}
\begin{aligned}
    \frac{d\mathcal{E}_{\gamma e^-\to e^- \overline{\chi}{\chi}}}{dVdt}=&\frac{4\alpha^3 q^2}{3(2\pi)^4}\int p_1 dp_1 q dq dx f_{e^-}(p_1) f_\gamma(q)\log\left[\frac{T^2}{m_e^2}\right]\sqrt{\frac{p_1 q(1-x)}{2}}\\ &\left[\frac{(p_1+q)\Psi_0\left(\frac{\sqrt{2}m_\chi}{\sqrt{p_1 q(1-x)}}\right)+\sqrt{p_1^2+q^2+2p_1 q x}\Psi_1\left(\frac{\sqrt{2}m_\chi}{\sqrt{p_1 q(1-x)}}\right)}{\sqrt{2p_1 q(1-x)}}\right]\,,
\end{aligned}
\end{equation}
where $x$ is the cosine of the angle between $\bp_1$ and $\bq$. In the logarithmic term, we have here replaced $\ttq\to T$ for consistency, since this term has been determined only with logarithmic precision. After rescaling all the variables by a factor $T$ to make them dimensionless, we finally find 
\begin{equation}
    \frac{d\mathcal{E}_{\gamma e^-\to e^- \overline{\chi}{\chi}}}{dVdt}=\frac{4\alpha^3 q^2 T^5 e^{\mu_e/T}}{3(2\pi)^4}\log\left[\frac{T^2}
    {m_e^2}\right]\Phi\left(\frac{m_\chi}{T}\right),
\end{equation}
with
\begin{equation}
\begin{aligned}
    \Phi(m)=&\int p dp q dq dx \frac{e^{-p}}{e^q-1}\sqrt{\frac{p_1 q(1-x)}{2}}\frac{(p+q)\Psi_0\left(\frac{\sqrt{2}m}{\sqrt{p_1 q(1-x)}}\right)+\sqrt{p^2+q^2+2p q x}\Psi_1\left(\frac{\sqrt{2}m}{\sqrt{p_1 q(1-x)}}\right)}{\sqrt{2p q(1-x)}}\\
    =& \int_0^\infty dq \int_{m^2/q}^\infty dp\int_{-1}^{1-\frac{2m^2}{p\,q}} dx \,p \, q  \frac{e^{-p}}{e^q-1}\frac{(p+q)\Psi_0\left(\frac{\sqrt{2}m}{\sqrt{p_1 q(1-x)}}\right)+\sqrt{p^2+q^2+2p q x}\Psi_1\left(\frac{\sqrt{2}m}{\sqrt{p_1 q(1-x)}}\right)}{2}\,,
\end{aligned}
\label{eq:URfunc}
\end{equation}
where the limits of integration are obtained by requiring that $\tilde{q}=\sqrt{\frac{p\,q\,(1-x)}{2}}>m_\chi$ and $-1\leq x\leq1$.
\end{widetext}

A good fit for $\Phi(y)$ in the range $10^{-4}\lesssim y \lesssim 10$ (with an accuracy better than $5\%$) is given by 
\begin{equation}
    \Phi (y) = \exp[2.97-1.44\,y^{1.08}-0.13 \log(y)]\,.
\end{equation}
In the low mass limit, the function is described in an excellent way (with precision better than $1\%$) by
\begin{equation}
    \Phi(y) = 11.88 -5.18 \log (y)\,\quad \quad y\lesssim 0.1
\end{equation}

The final result, after expressing the chemical potential $\mu_e$ in terms of the number density, is
\begin{equation}
\begin{aligned}
    \frac{d\mathcal{E}_{\gamma e^-\to e^- \overline{\chi}{\chi}}}{dVdt}&=\frac{2\pi^2\alpha^3 q^2 n_e T^2}{3(2\pi)^4}\log\left[\frac{T^2}{m_e^2}\right]\Phi\left(\frac{m_\chi}{T}\right)\\
    &\equiv q^2 n_e f_{UR}(m_\chi,T)\,,
\end{aligned}
\label{eq:Compt_UR}
\end{equation}
where we have defined
\begin{equation}
  f_{UR}(m_\chi,T) =  \frac{2\pi^2\alpha^3  T^2}{3(2\pi)^4}\log\left[\frac{T^2}{m_e^2}\right]\Phi\left(\frac{m_\chi}{T}\right)\,.
\label{eq:fUR}
\end{equation}
We stress that this result is valid only for $T>m_e$. The dotted lines in the left panel of Fig.~\ref{fig:EpsComp} show the Compton emissivity in the UR limit as a function of the temperature for $T>m_e$. The right panel display the relative error of the fit to $\Phi$ (red curve) as a function of $m_\chi/T$.

To find an expression valid for any value of the temperature, we interpolate between the NR and the UR regimes through a temperature-dependent interpolating function
\begin{equation}
    F(m_\chi,T) = \frac{f_{NR}(m_\chi,T)}{1+(T/m_e)^4}+\frac{f_{UR}(m_\chi,T)\, (T/m_e)^4}{1+(T/m_e)^4}\,,
\label{eq:Fint}
\end{equation}
where $f_{NR}$ and $f_{UR}$ are defined in Eq.~\eqref{eq:fNR} and Eq.~\eqref{eq:fUR}, respectively. We adopt the $(T/m_e)^4$ dependence so that the correct scaling of the emissivities is recovered in the NR and UR limits and the transition around $T\sim m_e$ is smooth. Finally, we define the Compton emissivity as
\begin{equation}
      \frac{d\mathcal{E}_{\gamma e^-\to e^- \overline{\chi}{\chi}}}{dVdt} = q^2 n_e F(m_\chi,T)  .
\label{eq:Compt_int}
\end{equation}
The solid lines in the left panel of Fig.~\ref{fig:EpsComp} represent the Compton emissivity as a function of the temperature $T$, obtained by interpolating between the NR (dashed) and UR (dotted) limits for three different MCP masses: $m_\chi = 30\,\rm{keV}$ (blue), $100\,\rm{keV}$ (red), and $300\,\rm{keV}$ (green).

\section{Pair annihilation}
\label{sec:PairProd}

\begin{figure*}
    \centering
    \includegraphics[width=0.48\textwidth]{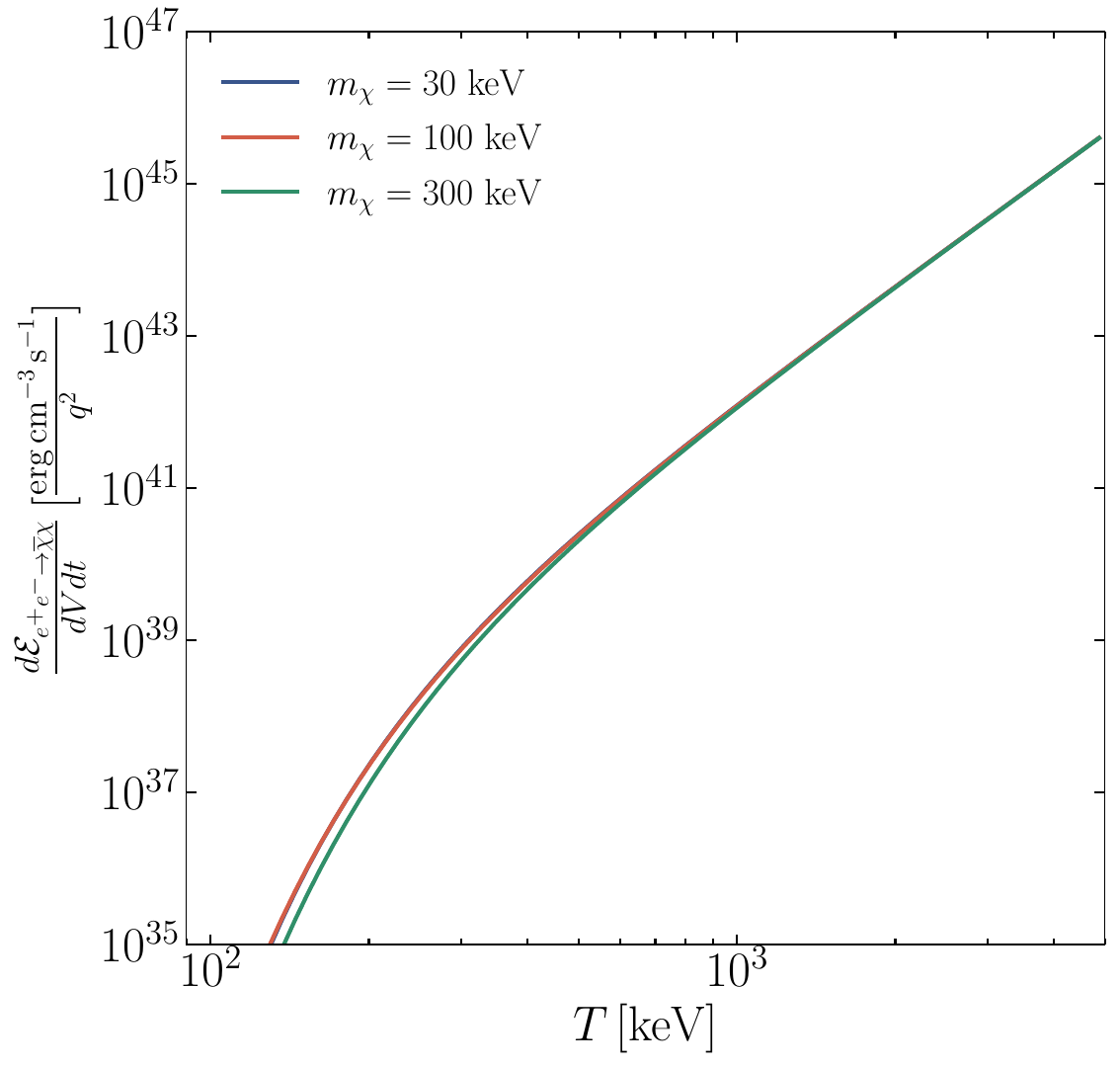}
    \includegraphics[width=0.49\textwidth]{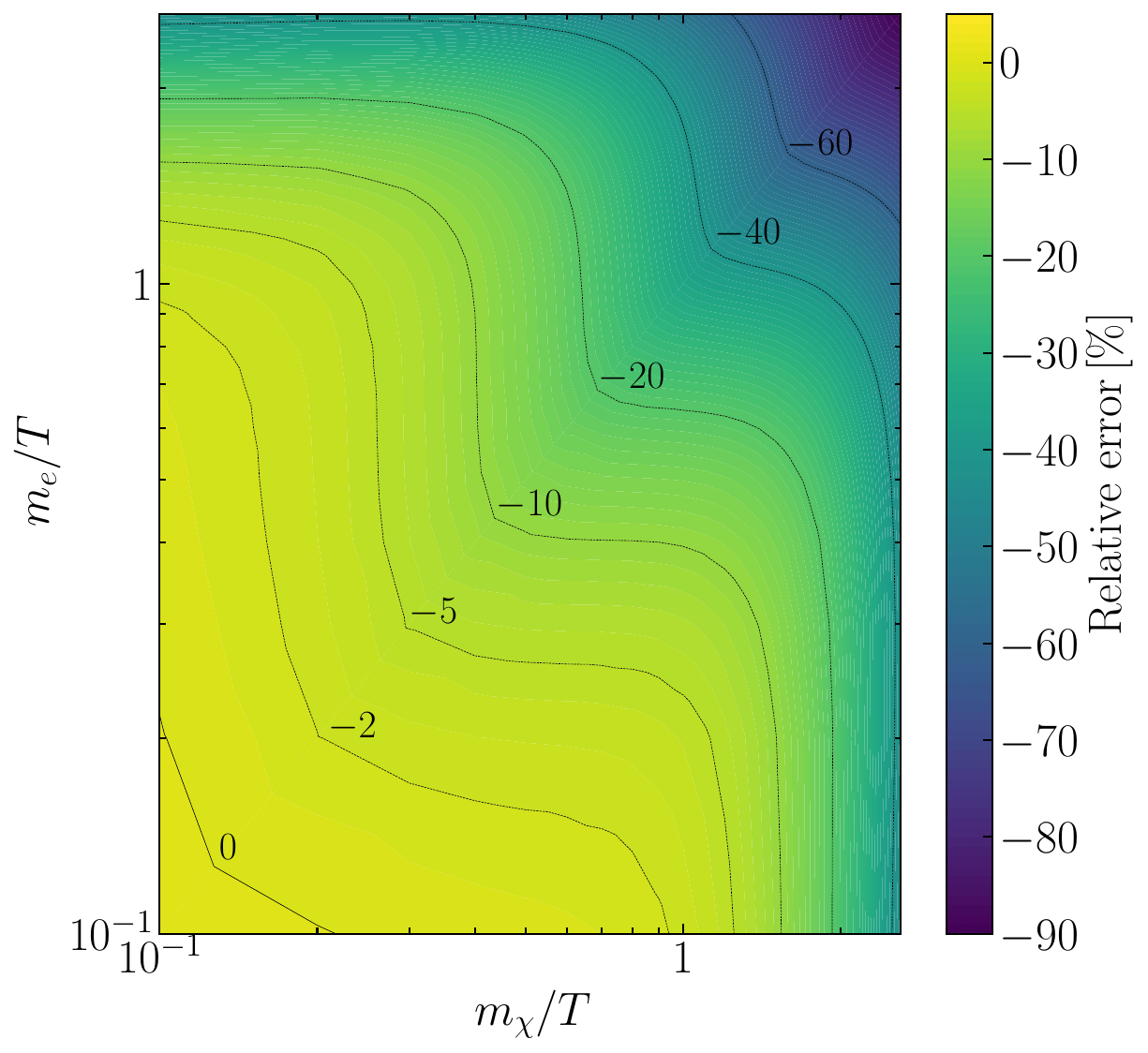}
    \caption{Left panel: MCP emissivity via pair production (see Eq.~\eqref{eq:em_pair}) as a function of the temperature $T$ for different values of MCP masses. Right panel: Relative error in the fit of the function $F$ defined in Eq.~\eqref{eq:F_pair}.}
    \label{fig:EpsPair}
\end{figure*}

Within an electron plasma, if the temperature is sufficiently high ($T\gtrsim m_e$), the amount of positrons may be large enough that pair annihilation $e^++e^-\to \overline{\chi}+\chi$ (the A process) becomes the dominant production channel. Under these conditions, the leptons in the plasma can be considered ultra-relativistic, since otherwise the amount of positrons would be strongly suppressed. In addition, as discussed in Sec.~\ref{sec:plasma}, the photons may also be taken to be ultra-relativistic, since $\omega_{\rm pl}\ll T$. For transverse photons, this means that we may consider them as massless photons, exactly as in a vacuum. For longitudinal plasmons, which exist only in a medium, this means that we may neglect them altogether, since a longitudinal massless electromagnetic field must decouple from any charge due to gauge invariance. 

The energy-loss rate per unit volume from the A process $e^++e^-\to \overline{\chi}+\chi$ can be computed from its total cross section, which in turn follows from the well-known cross section for $e^++e^-\to \mu^++\mu^-$ after rescaling by a factor $q^2$. In terms of the Mandelstam parameter for the scattering $s=2m_e^2+2\left[E_1 E_2-\bp_1\cdot \bp_2\right]$, where $\bp_1$, $\bp_2$ are the momenta of the two incoming electrons and $E_1$, $E_2$ their energies, we have
\begin{equation}
    \sigma_{\rm pair}=\frac{4\pi q^2 \alpha^2}{3s}\frac{(s+2m_e^2)(s+2m_\chi^2)}{s^2}\sqrt{\frac{s-4m_\chi^2}{s-4m_e^2}},
\end{equation}
where we have taken the limit $s\gg m_e^2$ since the reaction is kinematically allowed only for $s>4m_\chi^2\gg m_e^2$. In order to simplify this expression, we can take the limit $m_e\to 0$. In the case of massless MCP $m_\chi=0$, such an approximation would appear to lead to a potentially infinite cross section for collinear scattering $E_1 E_2=\bp_1\cdot\bp_2$. However, in reality this divergence is harmless since the relative velocity for collinear scattering vanishes. For $m_\chi\neq 0$, the Mandelstam parameter can never vanish, since it must satisfy $s>4m_\chi^2$, so that the argument of the square root is positive, implying that there is no divergence. Therefore, the limit $m_e\to 0$ does not entail any particular divergence, and is justified by the condition $T\gg m_e$ required for pair production to be a dominant process.

\begin{widetext}
The cooling rate per unit volume can now be written as
\begin{equation}
    \frac{d\mathcal{E}_{e^+ e^-\to \overline{\chi} \chi}}{dt dV}=\int \frac{2d^3\bp_1}{(2\pi)^3}\int \frac{2d^3\bp_2}{(2\pi)^3} (E_1+E_2)
    f_{e^-}(E_1) f_{e^+}(E_2) v_{\rm rel}\sigma_{\rm pair}.
\label{eq:em_pair}
\end{equation}
Here $f_{e^-}(E_1)$ and $f_{e^+}(E_2)$ are the electron and positron distribution functions and their product is, in the non-degenerate limit, simply $f_{e^-}(E_1) f_{e^+}(E_2)=e^{-\frac{E_1+E_2}{T}}$.
The relative velocity between the particles is
\begin{equation}
    v_{\rm rel}=\frac{\sqrt{s(s-4m_e^2)}}{2E_1 E_2}.
\end{equation}
The integral over the relative angle $\cos\theta_{12}$ between $\bp_1$ and $\bp_2$ can be expressed as an integral over the Mandelstam parameter.
Hence, introducing the dimensionless variables $\epsilon_1=E_1/T$ and $\epsilon_2=E_2/T$, the cooling rate may be expressed in terms of a single dimensionless function
\begin{equation}
     \frac{d\mathcal{E}_{e^+ e^-\to \overline{\chi} \chi}}{dt dV}=\frac{T^5q^2\alpha^2}{6\pi^3}F\left(\frac{m_e}{T},\frac{m_\chi}{T}\right),
\end{equation}
where
\begin{equation}
    F(x_e,x_\chi)=\int_{x_e}^{+\infty}d\epsilon_1\int_{x_e}^{+\infty}d\epsilon_2(\epsilon_1+\epsilon_2)e^{-\epsilon_1-\epsilon_2}\int_{2(\epsilon_1\epsilon_2+x_e^2-\sqrt{\epsilon_1^2-x_e^2}\sqrt{\epsilon_2^2-x_e^2})}^{2(\epsilon_1\epsilon_2+x_e^2+\sqrt{\epsilon_1^2-x_e^2}\sqrt{\epsilon_2^2-x_e^2})} d\hat{s}\frac{(\hat{s}+2x_e^2)(\hat{s}+2x_\chi^2)}{\hat{s}^2}\sqrt{1-\frac{4x_\chi^2}{\hat{s}}}.
\label{eq:F_pair}
\end{equation}

\end{widetext}
In the high-temperature limit the function is easily evaluated $F(0,0)=16$. Furthermore, while the expression of the function does not make it transparent, the function itself is symmetrical in its arguments $F(x_e,x_\chi)=F(x_\chi,x_e)$. We have found that a reasonable fit to the behavior of the function, in the range in which it is not too strongly suppressed, is
\begin{equation}
    F(x_e,x_{\chi})\approx 16.37\exp[-0.61 (x_e + x_\chi)^{1.73}+0.93 (x_e\,x_\chi)^{0.93}].
\end{equation}
We show in the left panel of Fig.~\ref{fig:EpsPair} the pair annihilation emissivity as a function of the temperature $T$ for different MCP masses. The right panel displays the relative error of the fit to $F(x_e,\,x_\chi)$, which remains below $20\%$ for temperatures larger than both the electron and MCP masses, where the production is not suppressed. At lower temperatures, the quality of the fit deteriorates; however, this is not phenomenologically relevant, as the process is strongly suppressed in that regime.

{For $T \gtrsim m_e$, the emissivity can be approximated to within $\sim 20\%$ by a simpler expression that depends only on $m_\chi$ and is obtained in the limit $m_e \to 0$:
\begin{equation}
\frac{d\mathcal{E}_{e^+ e^- \to \chi \bar{\chi}}}{dt\, dV}
= \frac{8 q^2 \alpha^2 T^5}{3\pi^3}
\exp\!\left[-0.718\,\left(\frac{m_\chi}{T}\right)^{1.332}\right]\,.
\end{equation}
}

\section{Discussion and Conclusions}\label{sec:disc}

\begin{figure*}[t!]
    \centering
    \includegraphics[width=1\columnwidth]{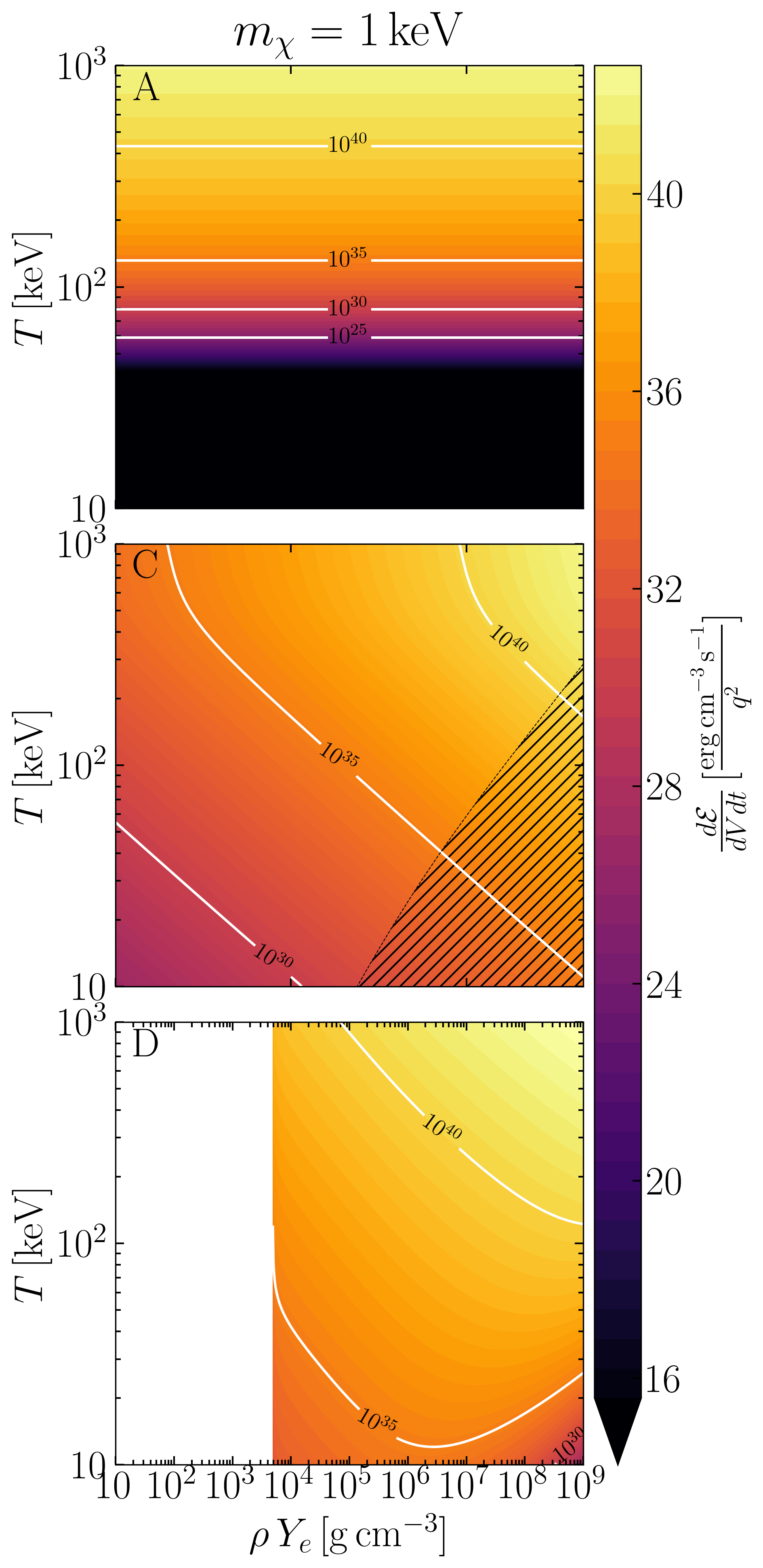}
        \includegraphics[width=1\columnwidth]{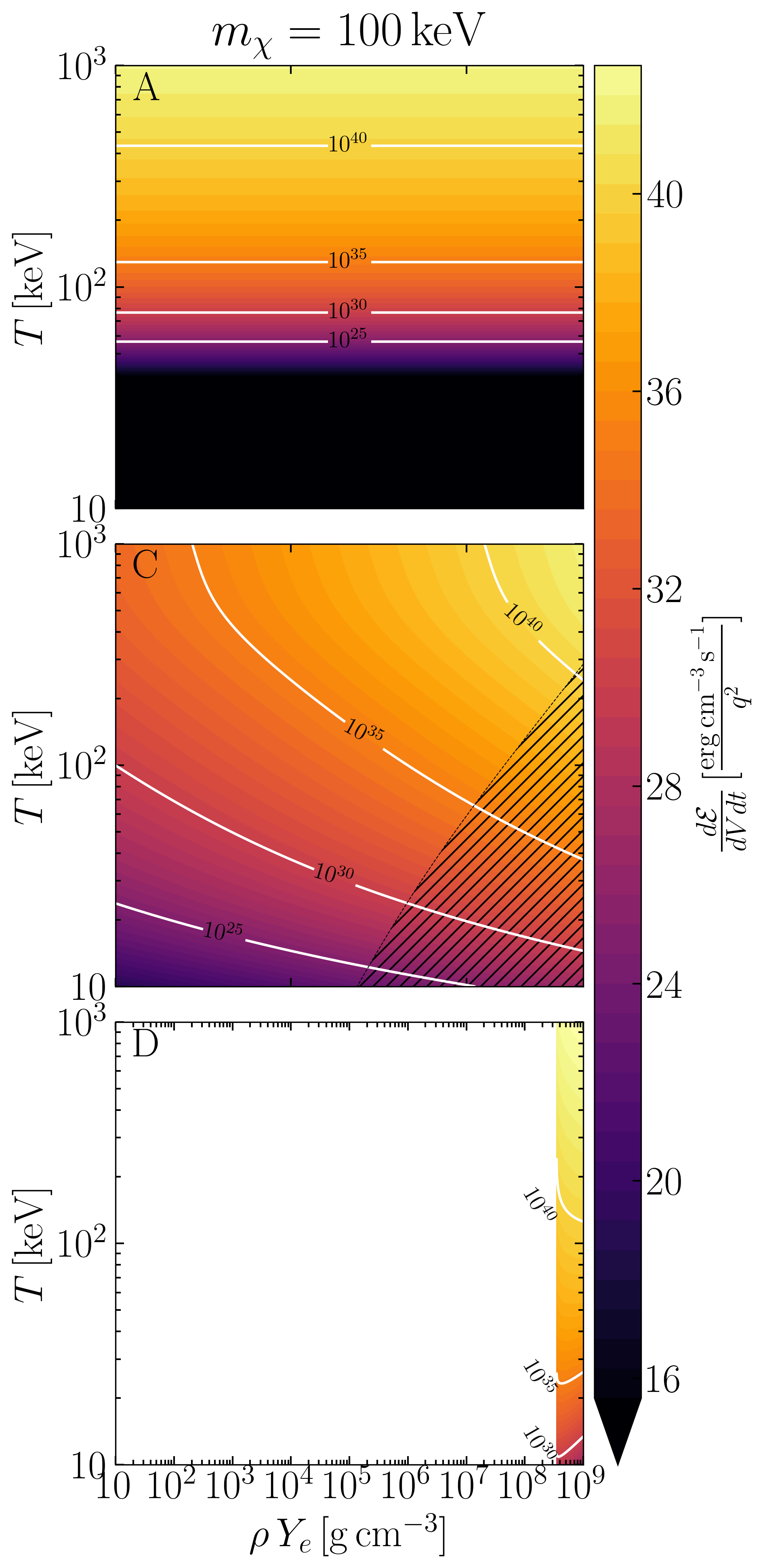}
\caption{Emissivities as a function of temperature and electron density for pair annihilation (A, top panels), Compton scattering (C, central panels), and plasmon decay (D, lower panels). Results are shown for $m_\chi = 1\,{\rm keV}$ (left panels) and $m_\chi = 100\,{\rm keV}$ (right panels). The black region in the upper panels corresponds to emissivities smaller than $10^{16}\,{\rm erg\,cm^{-3}\,s^{-1}}$. The hatched region in the central panels marks the conditions under which the plasma frequency cannot be neglected, since $\omega_{\rm pl}>T$. The white regions in the lower panels indicate where the D-process emissivity vanishes because $m_\chi > \omega_{\rm pl}/2$.}
    \label{fig:emiss}
\end{figure*}

\begin{figure*}[t!]
    \centering
    \includegraphics[width=1\columnwidth]{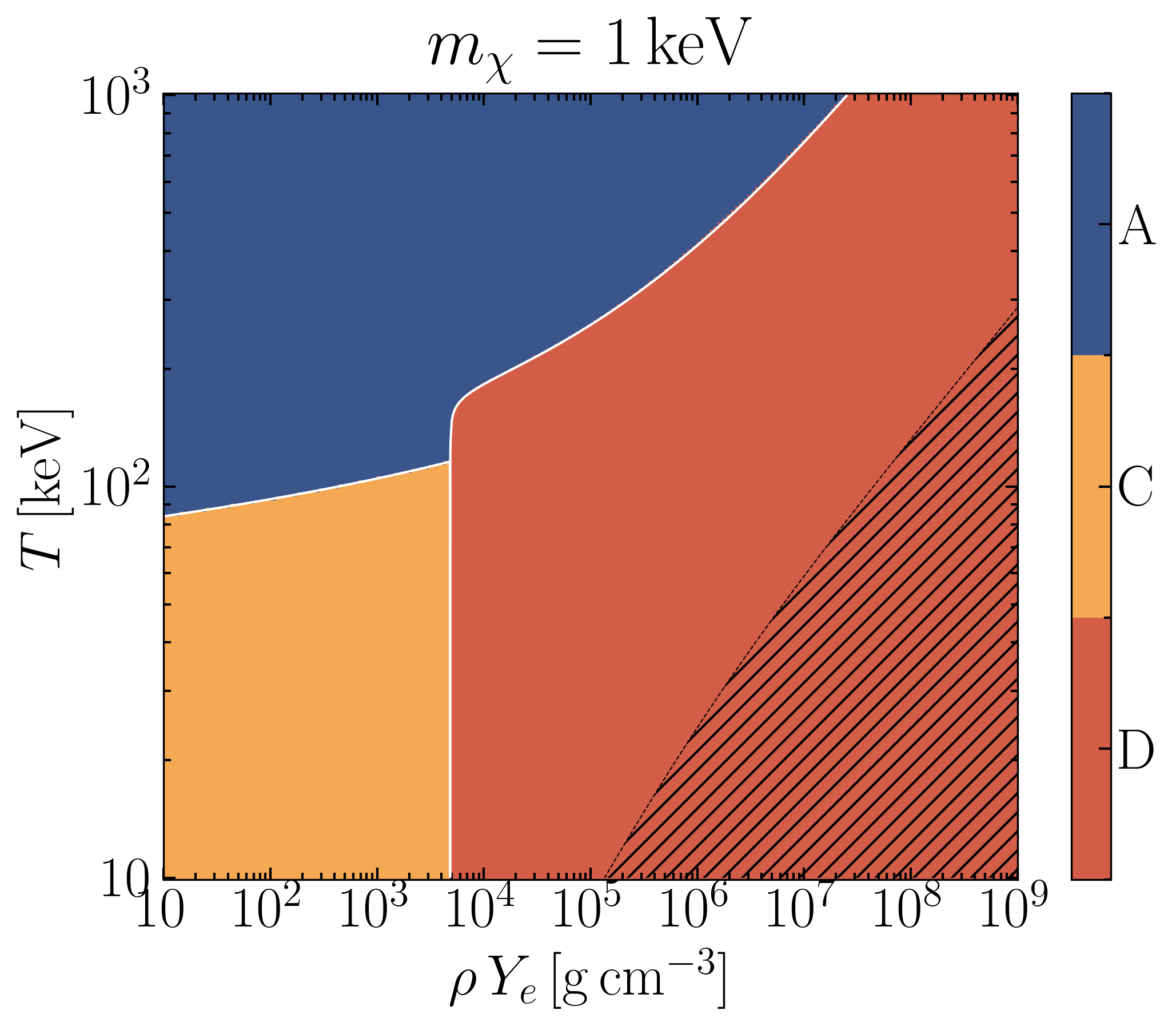}
      \includegraphics[width=1\columnwidth]{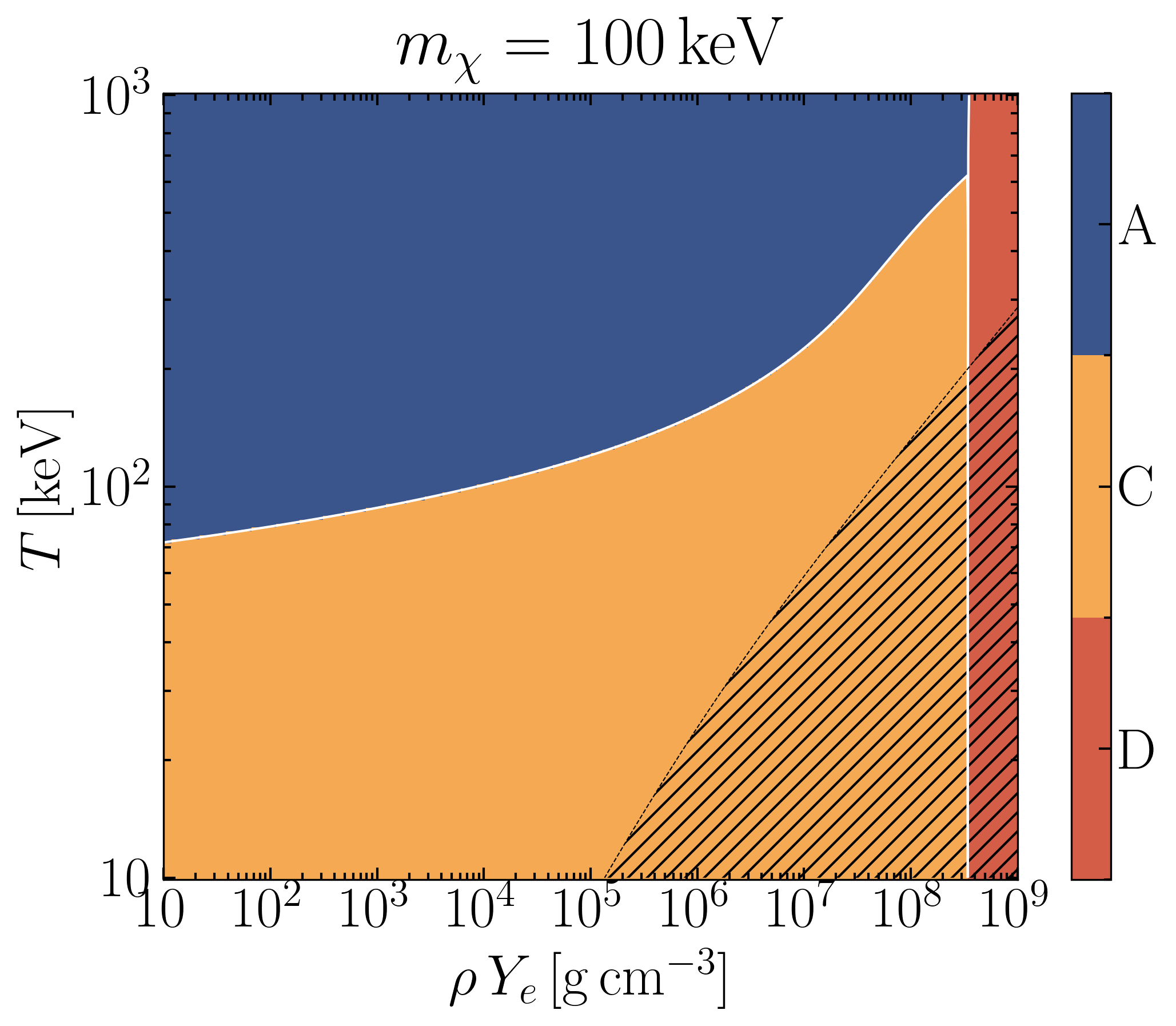}
    \caption{Dominant production MCP processes at different temperatures $T$ and electron densities $\rho\,Y_e$ for $m_\chi= 1\,{\rm keV}$ (left panel) and $m_\chi = 100\,{\rm keV}$ (right panel). The dominant process is determined by the largest emissivity among pair annihilation (blue), Compton (orange) and plasmon decay (red).
    }
    \label{fig:dominant_process}
\end{figure*}

Millicharged particles (MCPs) are featured in well-motivated extensions of the Standard Model. As their existence would have an impact on stellar evolution, relevant effort has been put into computing their production in stellar plasmas for different conditions. For the first time, we have computed the energy loss due to MCPs in the late stages of massive star evolution---before they undergo an implosion and become core-collapse supernovae.

We find that the dominant processes for MCP production depend mainly on the value of the temperature $T$, the plasma frequency $\omega_{\rm pl}$ and the MCP mass $m_\chi$. For typical stellar conditions, the relevant processes for MCP production are respectively:
plasmon {\bf D}ecay for $m_\chi < \omega_{\rm pl}/2$; {\bf C}ompton-like scattering for $m_\chi>\omega_{\rm pl}/2$ and $T\ll 500\,\rm{keV}$; pair {\bf A}nnihilation and {\bf C}ompton-like scattering for $m_\chi > \omega_{\rm pl}/2$ and $T\gtrsim500\,\rm{keV}$.

For masses $m_\chi < \omega_{\rm pl}/2$, the dominant production channel is transverse plasmon decay (D process), while the longitudinal contribution can be neglected. Throughout this work we assume $T\gg \omega_{\rm pl}$; otherwise, the opposite hierarchy would apply. The corresponding emissivity is given by
\begin{equation}
    \frac{d\mathcal{E}_{\gamma_T\to\chi\overline{\chi}}}{dVdt} = \frac{2\,q^2\,e^2 T^3}{3(2\pi)^3}\sqrt{1-\frac{4\,m_\chi^2}{\omega_{\rm pl}^2}}(\omega_{\rm pl}^2 + 2m_\chi^2) \Phi_T(\omega_{\rm pl}/T)\,,
\end{equation}
with
\begin{equation}
  \Phi_T(x) = \exp[0.889-0.346\,x^{1.314}]\,.
\end{equation}

For larger masses, the dominant production processes are the Compton effect (C process) if $T\ll m_e$, and the pair annihilation (A process) if $T\gtrsim m_e$. The Compton emissivity is given by
\begin{equation}
      \frac{d\mathcal{E}_{\gamma e^-\to e^- \overline{\chi}{\chi}}}{dVdt} = q^2 n_e F(m_\chi,T)  \,,
\end{equation}
where $F(m_\chi,T)$ is defined in Eq.~\eqref{eq:Fint}. This expression corrects the results of Refs.~\cite{Vogel:2013raa,Fung:2023euv}, especially for the longitudinal contribution, in the regime $T\ll m_e$, but most importantly is valid throughout the ultra-relativistic range with $T\gtrsim m_e$.

Finally, the emissivity via pair production is given by
\begin{equation}
     \frac{d\mathcal{E}_{e^+ e^-\to \chi \overline{\chi}}}{dt dV}=\frac{T^5q^2\alpha^2}{6\pi^3}F\left(\frac{m_e}{T},\frac{m_\chi}{T}\right),
\end{equation}
where $F(x_e,\,x_\chi)$ is defined in Eq.~\eqref{eq:F_pair}.

We show in Fig.~\ref{fig:emiss} the emissivity for the A (upper panels), C (central panels), and D (lower panels) processes as contour plots in the plane of electron density $\rho Y_e$ and temperature $T$. Two representative MCP masses are considered: a light MCP with $m_\chi = 1\,{\rm keV}$ (left panels), lower than the typical temperature in the late stages of massive stars, and a heavier MCP with $m_\chi = 100\,{\rm keV}$ (right panels). The plasma frequency is included only for the D process, while it is neglected for the Compton-like process. This approximation breaks down in the hatched region of the central panels, where $\omega_{\rm pl}>T$.

The A process is independent of the density, and its emissivity increases rapidly with temperature. It is strongly suppressed for $T<100\,{\rm keV}$, as indicated by the black region in the upper panels, corresponding to emissivities below $10^{16}\,{\rm erg\,cm^{-3}\,s^{-1}}$. The C process grows with both temperature and density. When the MCP mass exceeds the temperature, the emissivity becomes Boltzmann suppressed, as visible in the $T<100\,{\rm keV}$ region of the central right panel. Finally, the D process is kinematically allowed for $\rho Y_e \gtrsim 5\times10^{3}\,{\rm g\,cm^{-3}}$ when $m_\chi=1\,{\rm keV}$ and for $\rho Y_e \gtrsim 4\times10^{8}\,{\rm g\,cm^{-3}}$ when $m_\chi=100\,{\rm keV}$. Whenever it is kinematically accessible, the plasmon-decay emissivity exceeds the Compton contribution under the same thermodynamic conditions. In the regime $\omega_{\rm pl}\ll T$, the D-process emissivity increases with density, whereas at larger densities ($\omega_{\rm pl}>T$) it is suppressed due to the Boltzmann suppression of the plasmon population.

 \begin{figure*}[t!]
    \centering
        \includegraphics[width=0.49\textwidth]{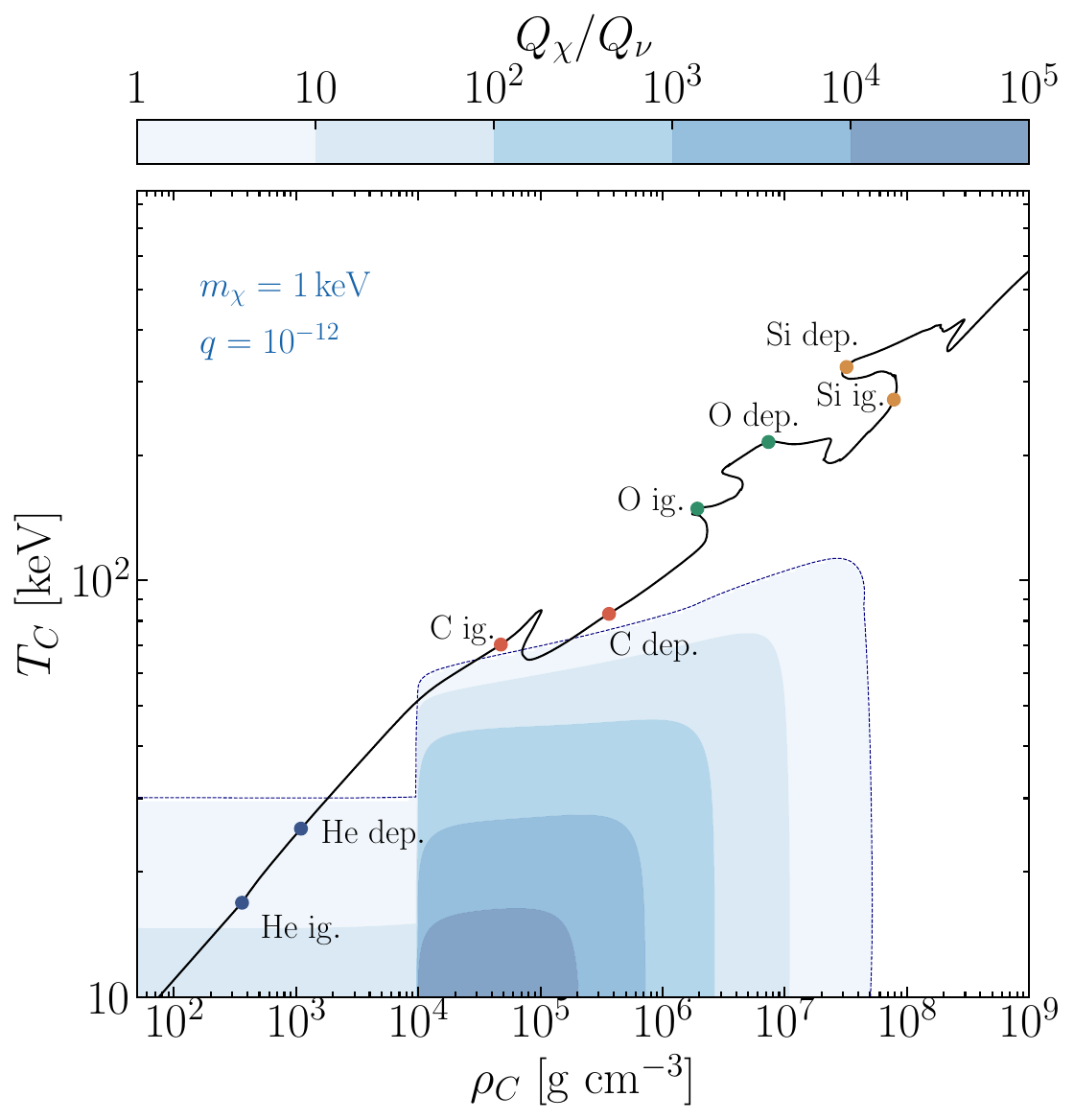}
    \includegraphics[width=0.49\textwidth]{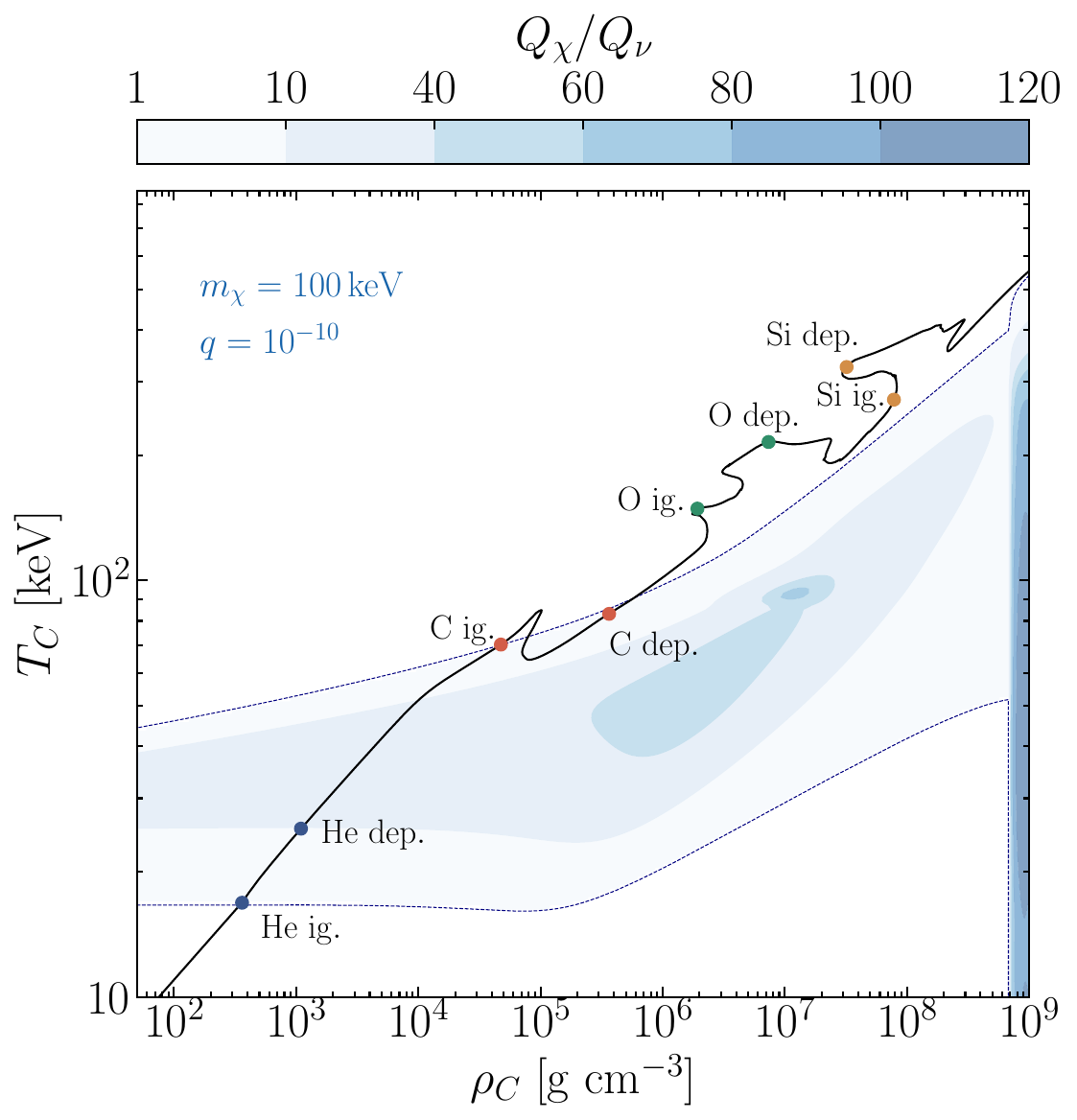}
    \caption{Evolutionary track of a $20\,M_\odot$ stellar model in the $T_C$–$\rho_C$ plane, overlaid with contours of the ratio $Q_\chi/Q_\nu$ for $m_\chi=1\,{\rm keV}$ and $q=10^{-12}$ (left panel), and for $m_\chi=100\,{\rm keV}$ and $q=10^{-10}$ (right panel). In both cases, $Q_\nu$ and $Q_\chi$ are computed assuming $Y_e=0.5$. Neutrino losses are taken from Ref.~\cite{1996ApJS..102..411I}, including the plasma, photoneutrino, and pair-annihilation processes relevant in massive stars.}
    \label{fig:ratioQchiQnu}
\end{figure*}

Figure~\ref{fig:dominant_process} shows plasma conditions under which the different processes dominate the MCP production for $m_\chi = 1\,{\rm keV}$ (left panel) and $m_\chi = 100\,{\rm keV}$ (right panel). For the lighter MCP mass, the C process dominates only when plasmon decay is kinematically forbidden (i.e., for $\rho\,Y_e \lesssim 5\times10^{3}\,{\rm g\,cm^{-3}}$) and $T \lesssim 100\,{\rm keV}$. At larger densities the D process becomes increasingly important, dominating for $T \lesssim m_e$ at $\rho\,Y_e \approx 10^{6}\,{\rm g\,cm^{-3}}$, and for $T \lesssim 10^{3}\,{\rm keV}$ at $\rho\,Y_e \approx 2\times10^{7}\,{\rm g\,cm^{-3}}$. For $m_\chi = 100\,{\rm keV}$ and $\omega_{\rm pl}<2m_\chi$, the Compton process dominates at $T \lesssim 100\,{\rm keV}$ for $\rho\,Y_e \lesssim 10^{5}\,{\rm g\,cm^{-3}}$, and at $T \lesssim m_e$ for $\rho\,Y_e \approx 10^{8}\,{\rm g\,cm^{-3}}$. At higher temperatures the A process becomes dominant. Once the plasmon decay channel is kinematically allowed, the D process dominates throughout the considered temperature range $10\,{\rm keV} < T < 10^{3}\,{\rm keV}$. 

Overall, for both light and heavier MCP masses the pattern of dominant production channels qualitatively resembles that of standard neutrino emission processes (see, e.g., Ref.~\cite{Haft:1993jt}). Finally, in the hatched region where $\omega_{\rm pl}>T$ and the plasma frequency cannot be neglected, bremsstrahlung---omitted here because it is subdominant in massive stars---may become relevant.

The sum of the contributions from the different processes gives the total MCP emissivity,
\begin{equation}
Q_\chi =
\frac{d\mathcal{E}_{\gamma_T\to\chi\overline{\chi}}}{dVdt}\,
\theta(\omega_{\rm pl}-2m_\chi)
+ \frac{d\mathcal{E}_{\gamma e^-\to e^- \chi \overline{\chi}}}{dVdt}
+ \frac{d\mathcal{E}_{e^+ e^-\to \chi \overline{\chi}}}{dVdt}\,,
\end{equation}
where the $\theta$ function accounts for the fact that plasmon decay is kinematically allowed only when $\omega_{\rm pl}>2m_\chi$. The Compton contribution is relevant only when plasmon decay is kinematically forbidden.

The presence of MCPs can affect stellar evolution if their emissivity $Q_\chi$ becomes comparable to or exceeds the total neutrino emissivity $Q_\nu$. In the late stages of massive stars, neutrinos are predominantly produced through plasma processes $\gamma_T\to\nu\overline{\nu}$, the photoneutrino process $\gamma e^-\to e^- \nu\overline{\nu}$, and pair annihilation $e^+e^-\to\nu\overline{\nu}$ (see, e.g., Refs.~\cite{Haft:1993jt,Raffelt:1996wa}). These correspond to the analogues of the D, C, and A processes for MCP production.

In Fig.~\ref{fig:ratioQchiQnu} we show contours of the ratio $Q_\chi/Q_\nu$ in the $T_C$–$\rho_C$ plane, together with the evolutionary track of the $20\,M_\odot$ model shown in Fig.~\ref{fig:ev_track}. Results are displayed for $m_\chi = 1\,{\rm keV}$ and $q=10^{-12}$ (left panel), and for ${m_\chi = 100\,{\rm keV}}$ and $q=10^{-10}$ (right panel). The discontinuities in the contours at $\rho_C \approx 7\times10^{8}\,{\rm g\,cm^{-3}}$ (left panel) and $\rho_C \approx 10^{4}\,{\rm g\,cm^{-3}}$ (right panel) correspond to the onset of the condition $\omega_{\rm pl}>2m_\chi$, which allows MCP production via plasmon decay. The MCP emission exceeds neutrino emission only in the regions enclosed by the dashed blue lines.

Since neutrino losses increase dramatically after He depletion and C ignition, Ref.~\cite{Croon:2020oga} argued that new particle emission can significantly affect the evolution of massive stars if it dominates over neutrino losses during the He-burning phase. This suggests that MCPs with $m_\chi\sim100\,{\rm keV}$ and $q\sim10^{-10}$---a region of parameter space that has not previously been constrained by astrophysical observations---may be probed through their impact on the evolution of massive stars.

Our semi-analytical fits for the energy-loss rates can be easily implemented in stellar evolution codes.
In a series of forthcoming papers, we will study the effect of MCPs on the late stages of stellar evolution~\cite{PPISN,R2}. 

\section*{Acknowledgments}
This article is
based on work from COST Action COSMIC WISPers
(CA21106), supported by COST (European Cooperation
in Science and Technology).
D.F.G.F. acknowledges support by the TAsP (Theoretical Astroparticle Physics) project, and was supported by the Alexander von Humboldt Foundation (Germany) for most of the completion of the project.
G.L. acknowledges support from the U.S. Department of Energy under contract number DE-AC02-76SF00515.~J.S.~is supported by NSF Grant No.~2207880.~E.V. acknowledges support from the Italian Ministero dell'Università e della Rircerca through the FIS 2 project FIS-2023-01577 (DD n. 23314 10-12-2024, CUP C53C24001460001) and through Departments of Excellence grant 2023--2027 ``Quantum Frontier'', as well as from Istituto Nazionale di Fisica Nucleare (INFN) through the Theoretical Astroparticle Physics (TAsP) project.

\bibliographystyle{bibi}
\bibliography{References}

\end{document}